\documentclass[a4paper, draftcls, onecolumn]{IEEEtran}

\ifCLASSOPTIONcompsoc
\else
\fi
\ifCLASSINFOpdf
\else
\fi
\usepackage{latexsym}
\usepackage[fleqn]{amsmath}
\usepackage{amssymb}
\usepackage{graphicx}

\newtheorem{theorem}{Theorem}

\newtheorem{algorithm}{Algorithm}

\newtheorem{remark}{Remark}
\newtheorem{definition}{Definition}
\newtheorem{procedure}{Procedure}
\newtheorem{condition}{Condition}

\newtheorem{ip}{IP Problem}

\newcommand{\by}{\mbox{\boldmath $y$}}

\begin{document}
%
\title{Almost Instantaneous Fix-to-Variable Length Codes}
%
%
%
%

\author{Hirosuke Yamamoto,~\IEEEmembership{Fellow,~IEEE,}
       Masato Tsuchihashi,
      and Junya Honda,~\IEEEmembership{Member,~IEEE,}
\IEEEcompsocitemizethanks{\IEEEcompsocthanksitem H.~Yamamoto and J.~Honda are with the Department of Complex Science and Engineering, 
The University of Tokyo, Kashiwa-shi, Chiba 277-8561, Japan.
\protect\\
E-mail: hirosuke@ieee.org, honda@it.k.u-tokyo.ac.jp
\IEEEcompsocthanksitem M.~Tsuchihashi is with the Department of Mathematical Informatics, The University of Tokyo, Bunkyo-ku, Tokyo 113-8656, Japan.
\protect\\
E-mail: Masato\_Tsuchihashi@mist.i.u-tokyo.ac.jp
}
\thanks{This work was supported in part by JSPS KAKENHI Grant Number 24656240.
}}

\IEEEtitleabstractindextext{%
\begin{abstract}
We propose almost instantaneous fixed-to-variable-length (AIFV) codes such that 
two (resp.~$K-1$) code trees are used if code symbols are binary (resp.~$K$-ary for $K\geq 3$), and source symbols are assigned to incomplete internal nodes in addition to leaves. 
Although the AIFV codes are not instantaneous codes, they are devised such that the decoding delay is at most two bits (resp. one code symbol) in the case of binary (resp. $K$-ary) code alphabet.
The AIFV code can attain better average compression rate than the Huffman code at the expenses of a little decoding delay and a little large memory size to store multiple code trees. 
We also show for the binary and ternary AIFV codes that the optimal AIFV code can be obtained by solving 0-1 integer programming problems.
\end{abstract}

}

\maketitle

\IEEEdisplaynontitleabstractindextext

%
\IEEEpeerreviewmaketitle

\section{Introduction}\label{sec-1}
Lossless source codes  are classified into 
fixed-to-variable-length (FV) codes and variable-to-fixed-length (VF) codes,
which can be represented by code trees and parse trees, respectively.
It is well known that the Huffman coding \cite{Huffman} and Tunstall coding \cite{Tunstall} can attain the best compression rate
in FV codes and VF codes, respectively, for stationary memoryless sources if a single code tree or a single parse tree is used.
But, Yamamoto and Yokoo \cite{Y-Y-01} showed that the AIVF (almost instantaneous variable-to-fixed length) coding can attain better compression rate than the Tunstall coding. 
An AIVF code uses $|{\cal X}|-1$ parse trees for a source alphabet ${\cal X}$ and 
codewords are assigned to incomplete internal nodes in addition to leaves in each parse tree. Although instantaneous encoding is not possible since incomplete internal nodes are used for encoding, the AIVF code is devised such that the encoding delay is at most one source symbol, and hence the code is called {\em almost instantaneous}. 
Furthermore, Yoshida and Kida \cite{Y-K-10}\cite{Y-K-12} showed that any AIVF code can be encoded and decoded by a single virtual multiple parse tree and the total number of nodes can be considerably reduced by the integration.

In the case of FV codes, it is well known by Kraft and McMillan Theorems \cite{Kraft}\cite{McMillan}\cite{C-T-05} that
any uniquely decodable FV code must satisfy Kraft's inequality, and such a code can be realized by an instantaneous FV code, i.e., a prefix FV code. Hence, the Huffman code, which can attain the best compression rate in the class of instantaneous FV codes, is also the best code in the class of uniquely decodable FV codes. 
However, it is assumed implicitly in the above argument  
that the best code  in uniquely decodable FV codes can be constructed by a fixed set of codewords (in other words, a single fixed code tree) for stationary memoryless sources. 
But, this assumption is not correct generally. Actually,  Yamamoto and Wei \cite{Y-W-2013} showed that we can devise more efficient FV codes than Huffman codes if multiple code trees can be used in the same way as the AIVF codes, and they called such FV codes $K$-ary AIFV (almost instantaneous fixed-to-variable length) codes when the size of code alphabet is $K$.
The $K$-ary AIFV code requires $K-1$ code trees to realize that the decoding delay is at most one code symbol. Hence, in the binary case with $K=2$, multiple code trees cannot be realized. To overcome this defect, they also proposed the binary AIFV code such that the decoding delay is at most two bits.
Although they proposed a greedy algorithm to construct a good AIFV code for a given source in 
\cite{Y-W-2013}, it is complicated and the optimal AIFV code cannot always be derived. Furthermore,  only a sketch is described for the binary AIFV codes, which are important practically, although $K$-ary AIFV codes for $K\geq 3$ are treated relatively in detail.

In this paper, we refine the definition of the binary and $K$-ary AIFV codes. 
The binary (resp.~$K$-ary for $K\geq 3$) AIFV code uses two (resp.~$K-1$) code trees, 
in which source symbols are assigned to incomplete internal nodes in addition to leaves. 
Although the AIFV codes are not instantaneous codes, they are devised such that the decoding delay is at most two bits (resp.~one code symbol) in the binary (resp.~$K$-ary) case.
Furthermore, for the binary and ternary AIFV codes, we give an algorithm based on integer programing to derive the optimal AIFV code for a given source. 

In Section \ref{sec-2}, we show some simple examples of ternary AIFV codes, which can attain better compression rate than the ternary Huffman codes.
Then, after we give the formal definition of $K$-ary AIFV codes for $K\geq 3$, we derive the Kraft-like inequality for the AIFV code trees.
Binary AIFV codes are treated in Section \ref{sec-3}.
Furthermore, we show in Section \ref{sec-4} that the optimal AIFV codes can be derived by solving 0-1 integer programming problems for the binary and ternary AIFV codes.  Finally, the compression rates of the AIFV codes are compared numerically with the Huffman codes for several source distributions in Section \ref{sec-5}.

\section{$K$-ary AIFV codes for $K\geq 3$.}\label{sec-2}
\subsection{Examples of ternary AIFV codes} \label{sec-2-1}
\begin{figure}[b]%
  \begin{center}
   \includegraphics[height=3.5cm]{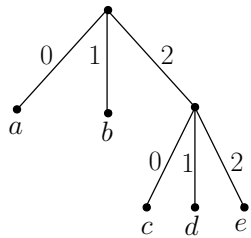} 
    \caption{The Ternary Huffman code for ${\cal X}=\{a, b, c, d, e\}$.}  \label{fig2}
  \end{center}
\end{figure}

We first consider a simple ternary FV code which encodes a source symbol  $x\in{\cal X}=\{a, b, c, d, e\}$ to a codeword in ${\cal Y}^*=\{0, 1,2\}^*$.
If the source distribution is uniform, i.e., $P(x)=1/5$ for all $x\in{\cal X}$, then the 
entropy of this source is $H_3(X) = \log_3 5\approx1.465$. The code tree of the Huffman code is given by Fig.~\ref{fig2} for this source, and the average code length $L_H$ of the Huffman code is $L_H=1.6$.

\begin{figure}[tb]%
  \begin{center}
  \includegraphics[height=3.5cm]{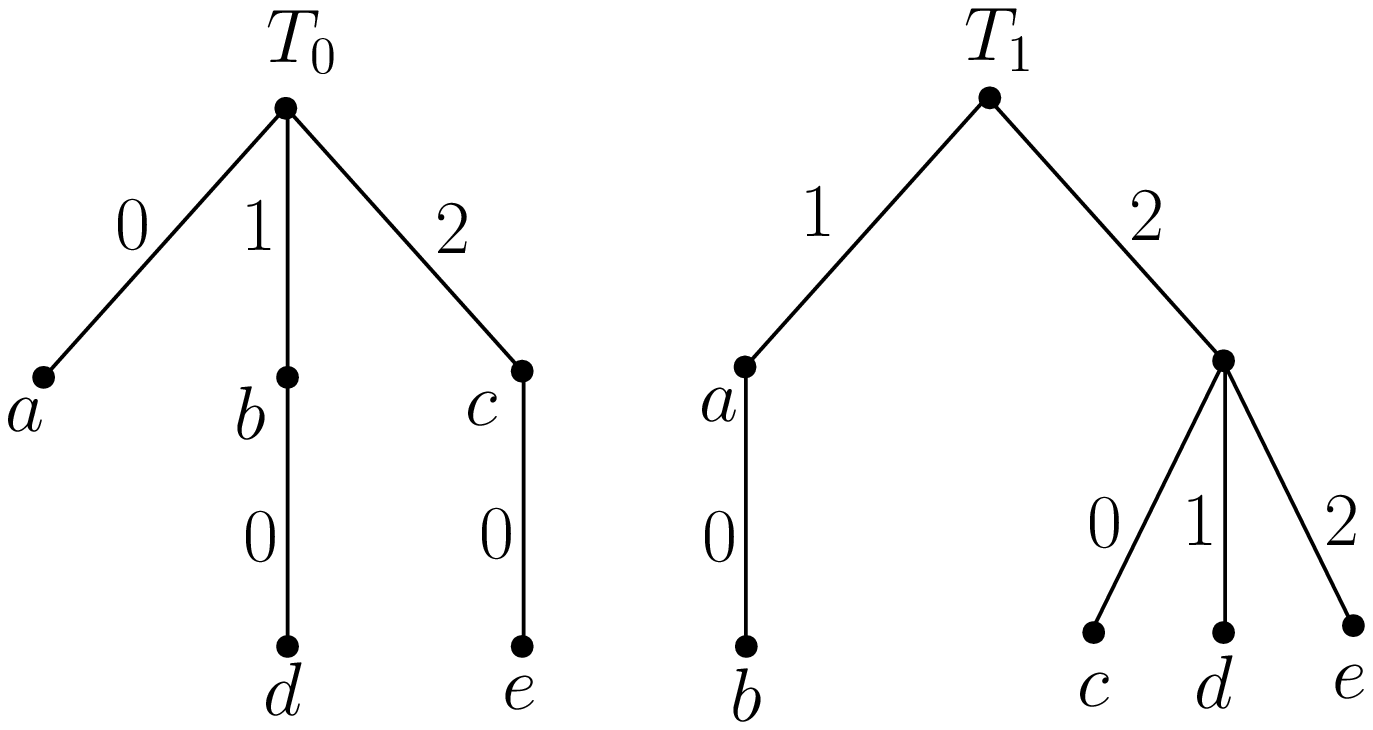}
  \caption{A ternary AIFV code for ${\cal X}=\{a, b, c, d, e\}$.}  \label{fig1}
  \end{center}
\end{figure}%

Next we consider a ternary AIFV code given by Fig.~\ref{fig1}, 
which satisfies the following properties.
\begin{definition}[Ternary AIFV codes]\label{3AIFV}
\begin{itemize}
\item[(A)] A ternary AIFV code consists of two code trees $T_0$ and $T_1$.
\item[(B)] Each complete internal node has three children connected by code symbols `0', `1', and `2', and each incomplete internal node has only one child connected by code symbol `0'
\footnote{
For simplicity, we say ``a node has a child connected by code symbol `$j$' " if the child is connected to the node by a branch with code symbol `$j$'.}.
\item[(C)] The root of $T_1$ must have two children connected by code symbols `1' and `2'.
\item[(D)] Source symbols are assigned to incomplete internal nodes in addition to leaves. But no source symbols are assigned to complete internal nodes. 
\end{itemize}
\end{definition}
The AIFV code encodes a source sequence $x_1x_2x_3\cdots$ as follows.
\begin{procedure}[Encoding of ternary AIFV codes] \label{e-3AIFV}
\begin{enumerate}      
\item[(a)] Use $T_0$ to encode the initial source symbol $x_1$.
\item[(b)] If $x_i$ is encoded by a leaf (resp.~an incomplete internal node),  then 
use $T_0$ (resp.~$T_1$) to encode the next source symbol $x_{i+1}$.
\end{enumerate}
\end{procedure}
When $T_0$ given by  Fig.~\ref{fig1} is used, the codewords of $a, b, c, d, e$ are 0, 1, 2, 10, 20, respectively. But, they are 1, 10, 20, 21, 22, respectively, when $T_1$ is used. 
For instance, source sequence `$abac$' is encoded to  `$0.1.1.20$' and
source sequence `$cdebac$' is encoded to `$2.21.20.1.1.20$', where dots `$.$' are 
inserted for the sake of human readability, but they are not necessary in the actual codeword sequences.

In the decoding of a codeword sequence $\by=y_1y_2y_3\cdots\in {\cal Y}^*$,  code trees $T_0$ and $T_1$ are used in the same way as the encoding.
\begin{procedure}[Decoding of ternary AIFV codes] \label{d-3AIFV}
\begin{enumerate}
\item[(a)] Use $T_0$ to decode the initial source symbol $x_1$ from $\by$.
\item[(b)] 
Trace $\by$ as long as possible from the root in the current code tree.
Then, output the source symbol assigned to the reached incomplete internal node or leaf.
\item[(c)]
Remove the traced prefix of $\by$, and 
if the reached node is a leaf (resp.~an incomplete internal node),  then use $T_0$ (resp.~$T_1$) to decode the next source symbol.
\end{enumerate}
\end{procedure}

For instance, if $\by=10020$, then the decoded sequence is $dae$
because `10', `0', and `20' correspond to leaves $d$, $a$, and $e$, respectively,  in  $T_0$. But, if $\by=1120$, 
$b$ is decoded from `1' in $T_0$ first because there is no path with `$11\cdots$' in $T_0$. Then, the current code tree transfers to $T_1$ because `1' corresponds to the incomplete internal node of $b$ in $T_0$.  By removing `1' from $\by$, we have $\by=120$. 
Next, $a$ is decoded from `1' in $T_1$ because there is no path with `$12\cdots$' in $T_1$, and the current code tree keeps $T_1$ because `1' corresponds to the incomplete internal node of $a$ in $T_1$. Finally, $c$ is decoded from `20' in $T_1$. 
Note that when a source symbol assigned to a leaf is decoded, the decoding is instantaneous.  On the other hand, the decoding is not instantaneous when a source symbol assigned to an incomplete internal node is decoded. But the decoding delay is only one code symbol even in such cases.

We now evaluate the average code length of the ternary AIFV code given by Fig.~\ref{fig1}.
Let $L_0$ and $L_1$ be the average code length of  $T_0$ and $T_1$, respectively. Then, we can easily show from Fig.~\ref{fig1} that $L_0=1.4$ and $L_1=1.8$.
The transition probability of code trees are given by 
\begin{align}
Q(T_1|T_0)&= P_X(b)+P_X(c)=0.4,\\
Q(T_0|T_1)&= P_X(b)+P_X(c)+P_X(d)+P_X(e)=0.8,
\end{align}
and the stationary probabilities of $T_0$ and $T_1$ are given by $Q(T_0)=2/3$ and $Q(T_1)= 1/3$, respectively. Hence, the average code length of the ternary AIFV code is given by
\begin{align}
L_{AIFV}=\frac{2}{3} L_0 + \frac{1}{3}L_1
      = \frac{4.6}{3} \approx 1.533, \label{eq-1}
\end{align}
which is shorter than the average code length of the Huffman code $L_H=1.6$.

Now we explain the reason why the AIFV code can beat the Huffman code.
Since incomplete internal nodes are used in addition to leaves for encoding in $T_0$, 
$L_0=1.4$ smaller than the source entropy $H_3(X)\approx 1.465$ can be realized. 
On the other hand,
$L_1=1.8$ is  larger than $L_H=1.6$ because the root of $T_1$ has only two children. But, from $Q(T_0) > Q(T_1)$, $L_{AIFV}$ is smaller than $L_H$ in the above example.

\begin{figure}[t]%
  \begin{center}
   \includegraphics[height=3.5cm]{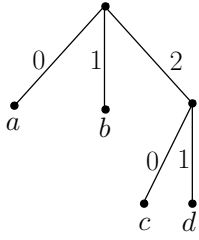} 
    \caption{The ternary Huffman code  for ${\cal X}=\{a, b, c, d\}$.}  \label{fig2-2}
  \end{center}
\end{figure}

\begin{figure}[tb]%
  \begin{center}
  \includegraphics[height=3.5cm]{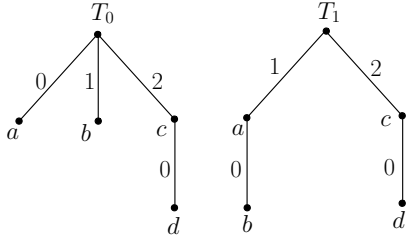}
  \caption{A ternary AIFV code for ${\cal X}=\{a, b, c, d\}$.}  \label{fig1-2}
  \end{center}
\end{figure}

If $|{\cal X}|$ is even, the loss of the ternary Huffman code becomes larger because the Huffman code tree must have an incomplete node. Consider the case that ${\cal X}=\{a, b, c, d\}$ and $P_X(x)=1/4$ for all $x\in {\cal X}$. Then the Huffman and AIFV code trees are given by Fig.~\ref{fig2-2} and Fig.~\ref{fig1-2}, respectively. In this case, the entropy of this source is $H_3(X)=\log_3 4\approx 1.262$, and the average code length is given by $L_H=1.5$ and $L_{AIFV}=4/3\approx1.333$. 

\begin{figure}[t]%
  \begin{center}
   \includegraphics[height=4cm]{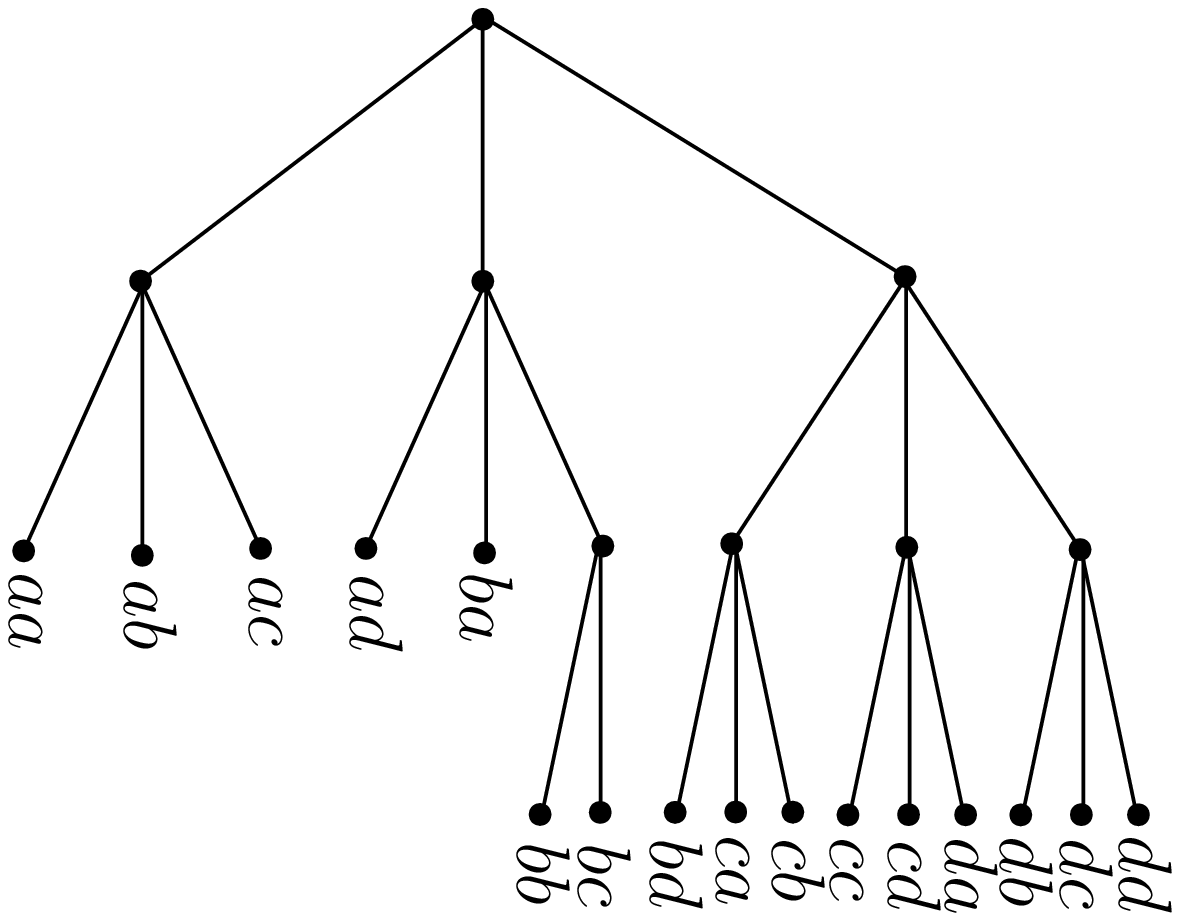} 
    \caption{A ternary Huffman code tree for ${\cal X}^2$.}  \label{fig2-3}
  \end{center}
\end{figure}

It is well known that if we construct the Huffman code for ${\cal X}^2$ as shown in Fig.~\ref{fig2-3}, the average code length per source symbol $L_H^{{\cal X}^2}$ can be improved compared with $L_H$. 
In the case of $P_X(x)=1/4$,  we have  $L_H^{{\cal X}^2}=43/32\approx 1.344 <L_H=1.5$.  But, the Huffman code for ${\cal X}^2$ has demerits such that
the size of the code tree increases to roughly $|{\cal X}|^2$, and the encoding and decoding delay becomes long as $|{\cal X}|$ becomes large.

\begin{figure}[t]%
  \begin{center}
   \includegraphics[height=4cm]{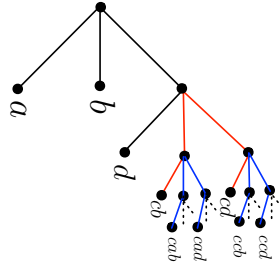} 
    \caption{The concatenated code tree of AIFV code.}  \label{fig3}
  \end{center}
\end{figure}

On the other hand, 
by concatenating $T_1$ to incomplete nodes of $T_0$ and $T_1$ in Fig.~\ref{fig1-2}, we obtain a code tree shown by Fig.~\ref{fig3}. Hence, the AIFV code can realize a flexible code tree for ${\cal X}^*$ by using only two code trees $T_0$ and $T_1$. We note that the total size of AIFV code trees is roughly $2|{\cal X}|$, and $L_{AIFV}=4/3\approx1.333$ is better than $L_H^{{\cal X}^2}=43/32\approx 1.344$. 
Furthermore, the encoding delay is zero and the decoding delay is at most one code symbol in the case of AIFV codes.

\begin{figure}[t]%
  \begin{center}
   \includegraphics[height=4cm]{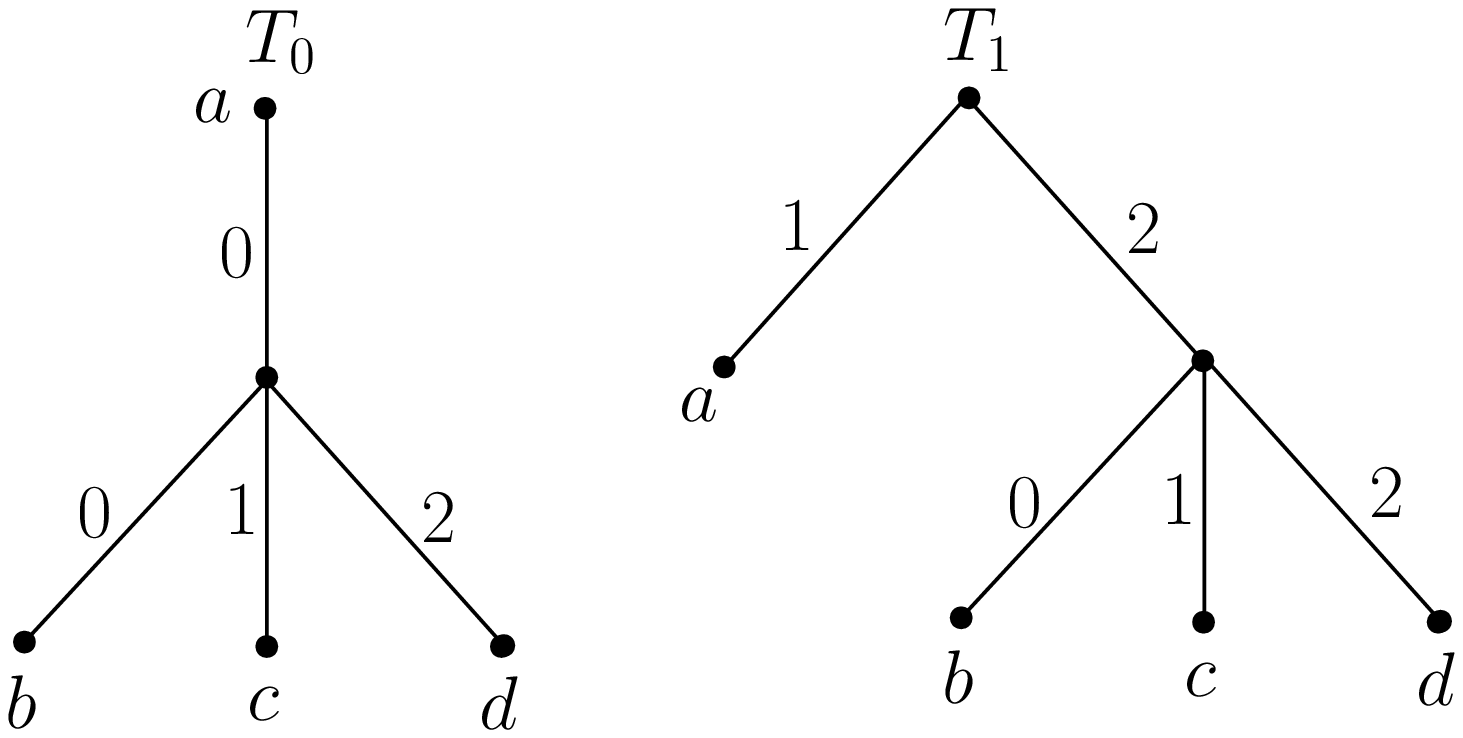} 
    \caption{A ternary AIFV code.}  \label{fig1-3}
  \end{center}
\end{figure}

We note from Definition \ref{3AIFV}  that the root of $T_1$ must have two children.
But, the root of $T_0$ can become an incomplete node.\footnote{The idea of assigning a source symbol to the root of $T_0$  was suggested by
Prof. M.~Nishiara at the presentation of \cite{Nishiara}.}
For instance, consider a source such that $P_X(a)=0.8$, $P_X(b)=0.1$, and $P_X(c)=P_X(d)=0.05$. In this case, the entropy is given by $H_3(X)\approx 0.6448$, and
the Huffman code of this source is given by Fig.~\ref{fig2-2} which attains $L_H(X)=1.1$. On the other hand, the ternary AIFV code shown in Fig.~\ref{fig1-3} attains $L_{AIFV}=Q(T_0)L_0+Q(T_1)L_1=(5/9)0.4+(4/9)1.2
\approx 0.7556$ for this source.
Note that $L_H$ cannot become shorter than 1 in any case while $L_{AIFV}$ can become shorter than 1 by assigning the source symbol $x$ with $P_X(x)\gg0.5$ to the root of $T_0$.
For instance, if we use the AIFV code shown in Fig.~\ref{fig1-3}, 
a source sequence $aabaaacd$ is encoded to `$\lambda.1.00.\lambda.1.\lambda.21.02$', where 
$\lambda$ represents the null codeword and dots `.' are not necessary in the actual codeword sequence. 
Hence the codeword sequence $\by$ is given by $\by=10012102$. 
Although the code length of $a$ in $T_0$ is zero, we can decode $x_1=a$ from the codeword sequence $\by$  because we begin the decoding with $T_0$, and there is no path with `$1\cdots$' in $T_0$, which means that $x_1$ is $a$.
Similarly we can decode `$aabaaacd$' from $\by=10012102$.
\footnote{Refer Remark \ref{remark-2} for how to detect the end of a source sequence. }

\subsection{$K$-ary AIFV codes}
In this subsection, we generalize ternary AIFV codes to $K$-ary AIFV codes with code alphabet ${\cal Y}=\{0, 1, 2, \cdots, K-1\}$ for $K\geq 3$.

\vspace{0.1cm}
\begin{definition}[$K$-ary AIFV code]\label{def-K-AIFV}\footnote{This definition is slightly different from \cite[Definition 1]{Y-W-2013} because the root of code tree $T_k$, $0\leq k\leq K-3$, can become incomplete in this paper although it must be complete in \cite[Definition 1]{Y-W-2013}.}
\label{KAIFV}
\begin{itemize}
\item[(A)] A $K$-ary AIFV code consists of $K-1$ code trees, $T_0, T_1, \cdots$, $T_{K-2}$.
\item[(B)]
Each complete internal node has $K$ children connected by code symbols
`0', `1', $\cdots$, `$K-1$'.  Every incomplete internal node has at least one and at most $K-2$ children connected by code symbols `0', `1', $\cdots$, `$K_c-1$',  where $K_c$ is the number of the children. 
\item[(C)]
The root of $T_k$ is called complete if it has $K-k$ children.
For $0\leq k\leq K-2$, the root of $T_k$ has $K-k$ children connected 
by code symbols `$k$', `$k+1$', $\cdots$ `$K-1$' if the root is complete.
For $0\leq k\leq K-3$, the root of $T_k$ can become incomplete,
and the incomplete root of $T_k$ must have at least one and at most $K-k-2$ children connected by `$k$', `$k+1$', $\cdots$, `$K_c-1$', where $K_c-k$ is the number of the children of the incomplete root. We regard the incomplete root of $T_k$ with $K_c-k$ children as an incomplete internal node with $K_c$ children.
\item[(D)]Source symbols are assigned to incomplete internal nodes in addition to leaves. But no source symbols are assigned to complete internal nodes. 
\end{itemize}
\end{definition}

\vspace{0.2cm}
A $K$-ary AIFV code can encode a source sequence $x_1x_2x_3\cdots$ 
and decode a codeword sequence $\by=y_1y_2y_3\cdots$
in the same way as ternary AIFV codes.

\newpage
\begin{procedure}[Encoding of $K$-ary AIFV codes] \label{e-KAIFV}
\begin{enumerate}
\item[(a)] Use $T_0$ to encode the initial source symbol $x_1$.
\item[(b)] When $x_i$ is encoded by a leaf (resp.~an incomplete internal node with $j$ children),  then use $T_0$ (resp.~$T_{j}$) to encode the next source symbol $x_{i+1}$.
\end{enumerate}
\end{procedure}

\begin{procedure}[Decoding of $K$-ary AIFV codes] \label{d-KAIFV}
\begin{enumerate}
\item[(a)] Use $T_0$ to decode the initial source symbol $x_1$ from $\by$.
\item[(b)] 
Trace $\by$ as long as possible from the root in the current code tree.
Then, output the source symbol assigned to the reached incomplete internal node or leaf.
\item[(c)]
Remove the traced prefix of $\by$, and 
if the reached node is a leaf (resp.~an incomplete internal node with $j$ children),  then use $T_0$ (resp.~$T_j$) to decode the next source symbol.
\end{enumerate}
\end{procedure}

\begin{figure}[t]%
  \begin{center}
   \includegraphics[height=5cm]{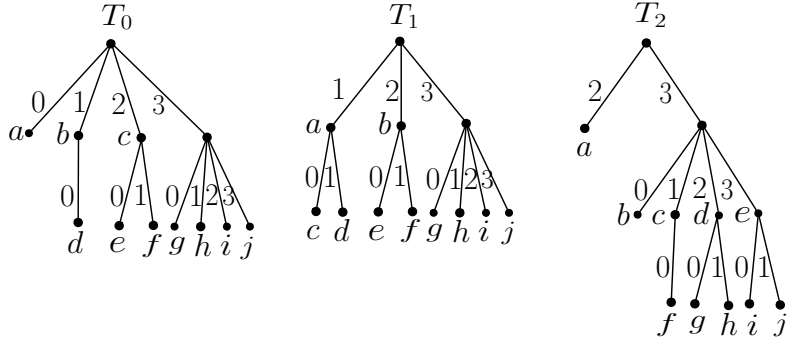} 
   \caption{An example of 4-ary AIFV code trees.}  \label{Exof4aryAIFV}
  \end{center}
\end{figure}

As an example, an AIFV code is shown in Fig.~\ref{Exof4aryAIFV} for the case of $K=4$ and 
${\cal X}=\{a, b, c, d, e, f, g, h, i, j\}$. When  source sequence
`$abacgcebbd$' is encoded by this AIFV code, the codeword sequence and the transition of code trees are given in Table \ref{T-4aryAIFV}.  
Note that when source symbol $x_i$ is encoded (or decoded) at a node with $j$ children,
then $x_{i+1}$ is encoded (or decoded) by $T_j$. 
Furthermore we can easily check that every $x_i$ can be uniquely decoded.
For instance, $x_2=b$ is encoded to codeword `1'  at incomplete internal node $b$ in $T_0$.
In this case, $x_3$ is encoded in $T_1$ because the incomplete internal node $b$ has one child in $T_0$.
This means that the codeword of $x_3$ does not begin with `0'. 
In the decoding, we obtain $\by=113130\cdots$ after the decoding of $x_1=a$ in $T_0$ and removing decoded codeword `$0$' from $\by$. 
Then we can decode $x_2=b$ because there is no path with $\by=11\cdots$  in $T_0$ but 
the path `$1$' corresponds to node $b$ in $T_0$.

\begin{figure}[t]%
  \begin{center}
   \includegraphics[height=5cm]{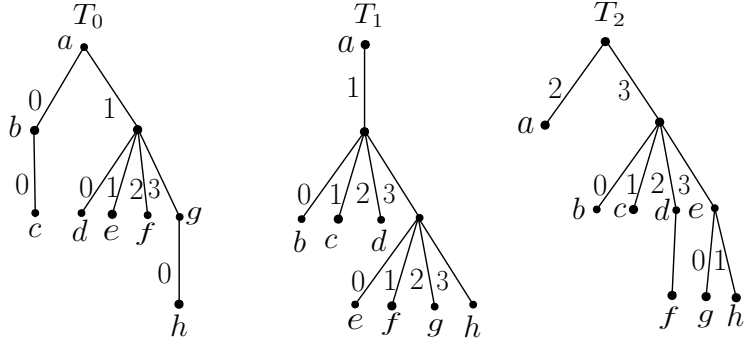} 
   \caption{An example of 4-ary AIFV code trees with incomplete roots.}  \label{Exof4aryAIFV-2}
  \end{center}
\end{figure}

Another example of 4-ary AIFV code trees for ${\cal X}=\{a, b, c, d, e, f, g, h\}$ is shown in Fig. ~\ref{Exof4aryAIFV-2}, in which the roots of $T_0$ and $T_1$ are incomplete.
The codeword sequence for  `$badbacgaec$' is shown in Table \ref{T-4aryAIFV-2},
where `$\lambda$' represents the null codeword. 
Note that the incomplete root of $T_k$ with $K_c-k$ children is regarded as an incomplete internal node with $K_c$ children as explained in Definition \ref{def-K-AIFV}-(C).
Hence, for instance,  node $x_2=a$ is the incomplete root with one child in $T_1$, and it is regarded as an incomplete internal node with 2 children. Hence $x_3$ is encoded (or decoded) in $T_2$.
In the decoding, we can decode $x_1=b$ from $\by=03210\cdots$ in $T_0$ because there is no path with $\by=03\cdots$ in  $T_0$, but path `0' corresponds to node $b$.
In the decoding of $x_2$, we have $\by=3210\cdots$ in $T_1$. But, there is no path which begins with `3'. Hence we obtain $x_2=a$ because `no path' means the root in $T_1$.

\begin{table}[t]
\begin{center}
\caption{An example of codeword sequence for 4-ary AIFV code.}\label{T-4aryAIFV}.
\begin{tabular}{c|cccccccccc}
$i$ & 1 &2& 3& 4& 5& 6& 7& 8& 9& 10\\
\hline
Current code tree& $T_0$ & $T_0$& $T_1$& $T_2 $& $T_1$&$T_0$&$T_2$&$T_2$&$T_0$&$T_1$\\
Source symbol $x_i$& $a$& $b$& $a$& $c$ & $g$ & $c$ & $e$ & $b$ & $b$ & $d$\\
Codeword &0& 1& 1& 31& 30& 2& 33& 30& 1& 11\\
Number of children of node $x_i$ & 0& 1& 2 &1&0& 2&2&0&1&0
\end{tabular}
\end{center}
\end{table}

Since some source symbols are assigned to incomplete internal nodes, the AIFV code is not 
an instantaneous code. But since the following theorem holds, this code is 
{\em almost instantaneously decodable}.
\begin{theorem}
The $K$-ary AIFV codes defined in  Definition \ref{def-K-AIFV} and Procedures \ref{e-KAIFV} and \ref{d-KAIFV} are uniquely decodable,  and the maximum decoding delay is at most one code symbol. 
\end{theorem}

{\em Proof:}
From Procedure \ref{e-KAIFV}-(b) and Procesure \ref{d-KAIFV}-(c), 
both encoding and decoding have the same transition of code trees. 
Hence, each source symbol $x_i$ is decoded in the same code tree used in the encoding.
It is clear from Procedure \ref{d-KAIFV}-(b) that if $x_i$ is encoded at a leaf in $T_k$, then 
$x_i$ is uniquely decodable. 
If $x_i$ is encoded at an incomplete internal node with $j$-children in $T_k$, then the children are connected 
by one of code symbols $\{0, 1, \cdots, j-1\}$ from the incomplete internal node.  
On the other hand, $x_{i+1}$ is encoded in $T_j$, in which any path begins with one of code symbols $\{j, j+1 \cdots, K-1\}$. Hence the node reached in  Procesure \ref{d-KAIFV}-(b) is the same incomplete internal node used in the encoding.

It is obvious that when $x_i$ is encoded at a leaf, then it can be decoded instantaneously. 
But, when $x_i$ is encoded at an incomplete internal node in $T_k$, we must read one more code symbol to judge  whether the incomplete internal node corresponds to the longest path in $T_k$. Hence the maximum decoding delay is at most one code symbol.

\begin{flushright} Q.E.D.\end{flushright}

\begin{table}
\begin{center}
\caption{An example of codeword sequence for 4-ary AIFV code with incomplete roots.}\label{T-4aryAIFV-2}.
\begin{tabular}{c|cccccccccc}
$i$ & 1 &2& 3& 4& 5& 6& 7& 8& 9& 10\\
\hline
Current code tree& $T_0$ & $T_1$& $T_2$& $T_1 $& $T_0$&$T_2$&$T_0$&$T_1$&$T_2$&$T_1$\\
Source symbol $x_i$& $b$& $a$& $d$& $b$ & $a$ & $c$ & $g$ & $a$ & $e$ & $c$\\
Codeword &0&$\lambda$& 32& 10& $\lambda$& 31& 13& $\lambda$& 33& 31\\
Number of children of node $x_i$ & 1& 1 &1&0& 2&0& 1&1&2&0
\end{tabular}
\end{center}
\end{table}

\vspace{0.2cm}
\begin{remark} \label{remark-1}
If there are no incomplete internal nodes with $j$ children in all code trees,  we can delete the code tree $T_{K-j}$. Furthermore, if we use only the incomplete internal nodes with $j$ children for a fixed $j$, $1\leq j \leq K-2$, then the code trees can be reduced to two code trees $T_0$ and $T_{K-j}$ 
even for the case of $K>3$. Such restriction worsens the compression rate of the $K$-ary AIFV codes. But, the construction of code trees becomes easy as shown in Section IV.
\end{remark}

\vspace{0.2cm}
\begin{remark} \label{remark-2}
In the decoding described in Procedures \ref{d-3AIFV} and \ref{d-KAIFV}, we assumed that the end of codeword sequence can be detected by another mechanism. In the case that the end cannot be detected and/or the null codeword is assigned to an incomplete root, we add a special symbol EOF to ${\cal X}$, and we assign EOF to a leaf in each $T_k$.  
By encoding EOF at the end of a source sequence, we can know the end of the decoding. 
The end of decoding can also be detected by adding the length of a source sequence
encoded by e.g.~Elias $\delta$ code \cite{Elias} into the prefix of the codeword sequence.
These worsen the compression rate a little. But, the degradation is negligible if $|{\cal X}|$ is not small and the length of a source sequence is sufficiently large.
\end{remark}

\subsection{Kraft-like inequalities for $K$-ary AIFV code trees}\label{Kraft-KAIFV}
In this subsection, we derive lower and upper bounds of average code length $L_k$ for code tree $T_k$, $0\leq k \leq K-2$. 

Let ${\cal N}_0^{(k)}$ (resp.~${\cal N}_j^{(k)}$) be the set of leaves (resp.~incomplete internal nodes with $j$ children) in code tree $T_k$, $0\leq k \leq K-2$, and 
let $n_x$ be the incomplete internal node or leaf corresponding to a source symbol $x\in {\cal X}$. Furthermore, let $l_k(x)$ be the code length of $x\in {\cal X}$ in  $T_k$.

We first consider $T_0$. If $n_x\in{\cal N}^{(0)}_j$, then we can change the node $n_x$ to a complete internal node by adding $K-j$ children at depth $l_0(x)+1$ of $T_0$.  
Hence, we have from Kraft's inequality that
\begin{equation}
\sum_{j=0}^{K-2} \sum_{x: n_x\in {\cal N}^{(0)}_j}(K-j) K^{-[l_0(x)+1]}  =1. \label{eq-10}
\end{equation}
In the case of $k>0$, since the root of $T_k$ has $K-k$ children, $K^{-[l_0(x)+1]}$ should become
$(K-k)^{-1}K^{-l_k(x)}$. Therefore, we have 
\begin{equation}
\sum_{j=0}^{K-2} \sum_{x: n_x\in {\cal N}^{(k)}_j} (K-j) (K-k)^{-1}K^{-l_k(x)} =1. \label{eq-11}
\end{equation}

Let $\hat{P}_X(x) = (K-j) (K-k)^{-1}K^{-l_k(x)}$ for $n_x\in {\cal N}^{(k)}_j$. 
Then, from $\sum_{x\in {\cal X}} \hat{P}_X(x)=1$
and $-\log_K \hat{P}_X(x) = l_k(x) + \log_K (K-k) - \log_K (K-j)$,
we have 
\begin{align}
0 &\leq D(P_X\|{\hat P}_X) 
= \sum_{x\in {\cal X}} P_X(x) \log_K \frac{P_X(x)}{\hat{P}_X(x)} \nonumber\\
& = -H_K(X) -\sum_{x\in {\cal X}} P_X(x) \log_K \hat{P}_X(x)  \nonumber\\
&= -H_K(X) + \sum_{j=0}^{K-2} \sum_{n_x\in {\cal N}^{(k)}_j} P_X(x) 
\left[ l_k(x) + \log_K (K-k) - \log_K (K-j)\right] \nonumber\\
 & =-H_K(X) + \left[\sum_{x\in {\cal X}} P_X(x)l_k(x)\right]  +\log_K (K-k) 
   -\sum_{j=0}^{K-2} \sum_{x: n_x\in {\cal N}^{(k)}_j} P_X(x)\log_K (K-j) \nonumber\\
 & =-H_K(X) + L_k + \log_K (K-k)
- \sum_{j=0}^{K-2} P({\cal N}^{(k)}_j) \log_K (K-j),\label{eq-12}
\end{align}
where $P({\cal N}^{(k)}_j)= \sum_{x: n_x\in {\cal N}^{(k)}_j}P_X(x)$. Hence, $L_k$ must satisfy that
\begin{equation}
L_k \geq H_K(X) +\sum_{j=0}^{K-2} P({\cal N}^{(k)}_j) \log_K \frac{K-j}{K-k}
\label{eq-13}
\end{equation}

Next we derive an upper bound of $L_k$.
If we allow that there exist leaves and/or incomplete internal nodes with no source symbol assigned in $T_k$,  \eqref{eq-11} becomes
\begin{equation}
\sum_{j=0}^{K-2} \sum_{x: n_x\in {\cal N}^{(k)}_j} (K-j) (K-k)^{-1}K^{-l_k(x)} \leq 1. \label{eq-14}
\end{equation}
Clearly, the original $T_k$ can attain better compression rate than such a relaxed code tree $\hat{T}_k$. 
We can easily check that $\hat{T}_k$ can be constructed if it satisfies \eqref{eq-14} 
and incomplete internal nodes can be arranged to satisfy the following condition.
\begin{condition}\footnote{Refer Section \ref{IP-tAIFV} to see how this condition can be represented by equations. }
 \label{cond-1}
Every node $n\in {\cal N}^{(k)}_j$ has $j$ children.
\end{condition}

We now define $l_k(x)$ as 
\begin{align}
l_k(x)& =\left\lceil -\log_K P_X(x) +\log_K \frac{K-j}{K-k}\right\rceil\nonumber\\
& < -\log_K P_X(x) +\log_K \frac{K-j}{K-k} +1. \label{eq-14-2}
\end{align}

Then, this $l_k(x)$ satisfies \eqref{eq-14}, and Condition \ref{cond-1} can 
be satisfied by setting $j$ appropriately for each $x$ because it can always be satisfied for any $x$ by $j=0$.
Hence, for appropriately selected $j$, 
we can construct $\hat{T}_k$ with average code length $\hat{L}_k$ satisfying that
\begin{align}
L_k \leq \hat{L}_k &=\sum_{x\in{\cal X}} P_X(x) l_k(x) \nonumber\\
&< H_K(X) +\sum_{j=0}^{K-2} P({\cal N}^{(k)}_j) \log_K \frac{K-j}{K-k} +1. \label{eq-15}
\end{align}
Note that the term $\log_K (K-j)/(K-k)$ in \eqref{eq-13} and \eqref{eq-15} is negative 
if $j> k$  although it is positive if $j< k$. Especially, in the case of  $L_0$,
the second term of \eqref{eq-13} and \eqref{eq-15} is always negative.

The global average code length $L_{AIFV}$ is given by
\begin{equation}
L_{AIFV} =\sum_{k=0}^{K-2} Q(T_k) L_k,  \label{eq-16}
\end{equation}
where $Q(T_k)$ is the stationary probability of $T_k$, and $Q(T_k)$ is determined from
$Q(T_j|T_k)=P({\cal N}^{(k)}_j)$, $0\leq k \leq K-2$, $0\leq j \leq K-2$.
Generally, it is difficult to evaluate the term in \eqref{eq-16} given by
\begin{align}
\sum_{k=0}^{K-2} Q(T_k) \sum_{j=0}^{K-2} P({\cal N}^{(k)}_j) \log_K \frac{K-j}{K-k} 
&= \sum_{k=0}^{K-2} \sum_{j=0}^{K-2} Q(T_k) Q(T_j|T_k)\log_K \frac{K-j}{K-k}.  
\label{eq-16-2}
\end{align}
But, in the case of $K=3$ or the case 
such that only two code trees are used for $K>3$ as described in Remark \ref{remark-2},
it holds that $Q(T_k) Q(T_j|T_k)=Q(T_j) Q(T_k|T_j)$. Hence, in these cases,
\eqref{eq-16-2} becomes zero, and the following bound is obtained from \eqref{eq-13} and \eqref{eq-15}--\eqref{eq-16-2}.
\begin{equation}
 H_K(X) \leq L_{AIFV} < H_K(X)+1  \label{eq-17}
\end{equation}

Unfortunately, the upper bound in \eqref{eq-17} is the same as the well known bound of the Huffman code. But, this fact does not mean that the performance of AIFV code with two code trees is the same as the performance of the Huffman code. 
The AIFV code trees are more flexible than the Huffman code tree. The term `$+1$' in \eqref{eq-14-2} can be decreased by selecting $j$ appropriately for each $x\in{\cal X}$ in the case of AIFV code trees.
Actually, as we will show in Section \ref{sec-6}, the AIFV codes can attain better compression rate than the Huffman codes.

\section{Binary AIFV codes}\label{sec-3}
\subsection{Definition of binary AIFV codes}
The $K$-ary AIFV codes treated in the previous section can be constructed only for $K\geq 3$, and the binary represented codewords of $K$-ary AIFV codes
are not so short as binary Huffman codes. But, we show in this section that if  decoding delay  is allowed at most two bits,
we can construct a binary AIFV code that attains better compression rate than the binary Huffman code. 

\begin{figure}[t]%
  \begin{center}
  \includegraphics[height=3.5cm]{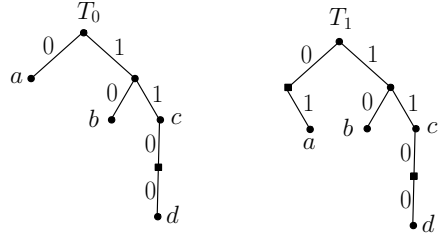}
  \end{center}
   \vspace*{-0.5cm} 
  \caption{A binary AIFV code.}
  \label{fig6}
  \vspace*{-0.2cm} 
\end{figure}

We first show a simple example of a binary AIFV code in Fig.~\ref{fig6}, which satisfies the following properties.

\begin{definition}[Binary AIFV codes]\label{2AIFV}
\begin{itemize}
\item[(A)] A binary AIFV code consists of two code trees $T_0$ and $T_1$.
\item[(B)] Each complete internal node has two children connected by code symbols `0', and `1'.  Incomplete internal nodes, each of which has one child, are divided into two categories, say master nodes and slave nodes. 
The child of a master node must be a slave node, and 
the master node is connected to its grandchild by code symbols `00'.
\item[(C)] The root of $T_1$ must have two children 
connected by code symbols `0' and `1'.
The child connected by `0' is a slave node and the root cannot have a grandchild connected by code symbols `00'.
\item[(D)] Source symbols are assigned to master nodes in addition to leaves. 
But no source symbols are assigned to neither complete internal nodes nor slave nodes.
\end{itemize}
\end{definition}
The binary AIFV code encodes a source sequence $x_1x_2x_3\cdots$ as follows.
\begin{procedure}[Encoding of binary AIFV codes] \label{e-2AIFV}
\begin{enumerate}
\item[(a)] Use $T_0$ to encode the initial source symbol $x_1$.
\item[(b)] When $x_i$ is encoded by a leaf (resp.~a master node),  then 
use $T_0$ (resp.~$T_1$) to encode the next source symbol $x_{i+1}$.
\end{enumerate}
\end{procedure}
If we use the binary AIFV code shown in Fig.~\ref{fig6}, then for instance, a source sequence 
`$cbcaab$' is encoded to  `$11.10.11.01.0.10$', and
source sequence `$cadbca$' is encoded to `$11.01.1100.10.11.01$', where dots `$.$' are not necessary in the actual codeword sequences.

A codeword sequence $\by=y_1y_2y_3\cdots\in {\cal Y}^*$ can be decoded by using  code trees $T_0$ and $T_1$ as follows.
\begin{procedure}[Decoding of binary AIFV codes] \label{d-2AIFV}
\begin{enumerate}
\item[(a)] Use $T_0$ to decode the initial source symbol $x_1$ from $\by$.
\item[(b)] 
Trace $\by$ as long as possible from the root in the current code tree.
Then, output the source symbol assigned to the reached master node or leaf.
\item[(c)]
Let $\hat{\by}$ be the path from the root to the reached master node or leaf.
Then, remove $\hat{\by}$ from the prefix of $\by$.
If the reached node is a leaf (resp.~a master node),  then use $T_0$ (resp.~$T_1$) to decode the next source symbol.
\end{enumerate}
\end{procedure}

For instance, from $\by=11101101010$, we can decode $x_1=c$  when `111' is read 
because there is no path  `111' from the root in $T_0$ but the  master node $c$ is reached by `11'. Similarly, in the case of $\by=11011100101101$, we can decode $x_1=c$ when `1101' is read because there is no path `1101' in $T_0$. 
We can easily check that `$cadbca$' can be decoded from $\by=11011100101101$.
We note that $x_i$ is decoded instantaneously if $x_i$ is encoded by a leaf,
and it is decoded with two-bit delay if $x_i$ is encoded by a master node.
Hence, the decoding delay of the binary AIFV codes is at most two bits.

Now consider a source such that 
${\cal X}=\{a, b, c, d\}$, and  $P_X(a)=0.45$, $P_X(b)=0.3$, $P_X(c)=0.2$,  $P_X(d)=0.05$.
In this case, the entropy and the average code length of the binary Huffman code 
are given by $H_2(X)\approx1.7200$ and $L_H=1.8$, respectively.
If we use the binary AIFV code shown in  Fig.~\ref{fig6}, the average code length are given by $L_0=1.65$ and $L_1=2.1$ for $T_0$ and $T_1$, respectively. Since $T_1$ is used only just after $c$ is encoded in Fig.~\ref{fig6},  we have $Q(T_1|T_0)=0.2$ and $Q(T_0|T_1)=0.8$ which mean that $Q(T_0)=0.8$ and $Q(T_1)=0.2$. Therefore, we have $L_{AIFV}=1.65\times 0.8+ 2.1\times 0.2 =1.74$, which is better than $L_H=1.8$.

\begin{figure}[t]%
  \begin{center}
  \includegraphics[height=3.5cm]{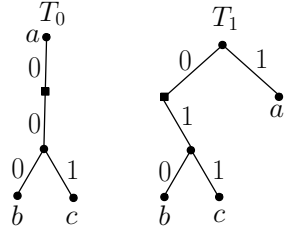}
  \end{center}
   \vspace*{-0.5cm} 
  \caption{A binary AIFV code with the incomplete root in $T_0$.}
  \label{2ary-2}
  \vspace*{-0.2cm} 
\end{figure}

Note that the root of $T_0$ can become a master node although the root of $T_1$ must have two children. Such an AIFV code is shown in Fig.~\ref{2ary-2} for ${\cal X}=\{a, b, c\}$.
For instance, source sequence $x_1x_2x_3=aaab$ is encoded to codeword sequence `$\lambda.1.\lambda.010$' by this AIFV code, which means $\by=1010$.
We can decode $x_1x_2x_3$ uniquely from $\by=1010$. First, we decode $x_1=a$ because there is no path with `$1\cdots$' in $T_0$. This means that $x_1$ is encoded at the root of $T_0$, and hence $x_1=a$. Next we move to $T_1$, and we obtain $x_2=a$ from $\by=1010$. Then, we move to $T_0 $ with $\by=010$. Since there is no path with `$1\cdots$' in $T_0$, we decode $x_3=a$. Finally we move to $T_1$ with $\by=010$, and we obtain $x_4=b$.
When $P_X(a)=0.9$ and $P_X(b)=P_X(c)=0.05$, this AIFV code have that $Q(T_1|T_0)=0.9$, 
$Q(T_0|T_1)=1$, $Q(T_0)=10/19$, $Q(T_1)=9/19$, $L_0=0.3$, $L_1=1.2$, and $L_{AIFV}=Q(T_0)L_0+Q(T_1)L_1\approx 0.7263$. On the other hand, this source has $H(X)\approx 0.5690$ 
and the average code length of the Huffman code  is $L_H=1.1$. 
In the binary case, $L_H$ cannot become shorter than one while $L_{AIFV}$ can become shorter than one as shown in this example.

\subsection{Kraft-like inequalities for binary AIFV codes}\label{Kraft-bAIFV}
In the same way as Section \ref{Kraft-KAIFV}, 
we can derive Kraft-like inequalities for binary AIFV codes.
Let ${\cal N}_0^{(k)}$ (resp.~${\cal N}_1^{(k)}$) be the set of leaves (resp. master nodes) in code tree $T_k$, $k=0, 1$. Furthermore, let $n_x$ be the master node or leaf assigned a source symbol $x$, and let $l_k(x)$ be the code length of $x\in {\cal X}$. 
Note that since a master node has only one grandchild, the master node becomes a complete node if we add three grandchildren to the master node. 
Hence we have the following relation for $T_0$. 
\begin{align}
\sum_{x: n_x\in {\cal N}_0^{(0)}} 2^{-l_0(x)} + \frac{3}{4}
 \sum_{x: n_x\in {\cal N}_1^{(0)}} 2^{-l_0(x)}=1 . \label{eq-20}
\end{align}
Similarly, the following relation holds for $T_1$ because the root of $T_1$ can have only three grandchildren.
\begin{align}
\sum_{x: n_x\in {\cal N}_0^{(1)}} 2^{-l_1(x)} + \frac{3}{4}
\sum_{x: n_x\in {\cal N}_1^{(1)}}  2^{-l_1(x)}=\frac{3}{4} \label{eq-20-1}
\end{align}
or 
\begin{align}
\frac{4}{3}\sum_{x: n_x\in {\cal N}_0^{(1)}} 2^{-l_1(x)} + \sum_{x: n_x\in {\cal N}_1^{(1)}}
 2^{-l_1(x)}=1 .\label{eq-20-2}
\end{align}
Furthermore, the global average code length $L_{AIFV}$ is given by
\begin{align}
L_{AIFV}&=Q(T_0)L_0+Q(T_1)L_1\nonumber\\
& =\frac{P({\cal N}_0^{(1)})L_0 +P({\cal N}_1^{(0)})L_1}{P({\cal N}_0^{(1)})+P({\cal N}_1^{(0)})}. \label{eq-20-3}
\end{align}

Then, in the same way as \eqref{eq-13}, \eqref{eq-15}, 
and \eqref{eq-17}, we can derive the following bounds.
\begin{align}
H_2(X) -P({\cal N}_1^{(0)})(2-\log_2 3)&\leq L_0 
< H_2(X) -P({\cal N}_1^{(0)})(2-\log_2 3) +1, \label{eq-21}\\
 H_2(X) +P({\cal N}_0^{(1)})(2-\log_2 3) &\leq L_1 
<  H_2(X) +P({\cal N}_0^{(1)})(2-\log_2 3) +1,  \label{eq-22}\\
 H_2(X) \leq L_{AIFV} <H_2(X)+1,  \label{eq-23}
\end{align}
where the upper bounds of the above inequalities must satisfy the following condition.
\begin{condition}\footnote{Refer Section \ref{IP-bAIFV} to see how this condition can be represented by equations. }
 \label{cond-2}
Every node $n\in {\cal N}^{(k)}_1$, $k=0,1$, has one grandchild.
\end{condition}

Note that $L_0$ can become smaller than the source entropy $H_2(X)$  but $L_1$ is larger than $H_2(X)$. 
Although the upper bound $H_2(X)+$1 in  \eqref{eq-23} is the same as the case of Huffman codes, the term `+1' can be decreased than the Huffman codes for individual sources because the binary AIFV code trees are more flexible than the Huffman code tree.

\section{Construction of AIFV code trees based on integer programming}\label{sec-4}
In this section, we propose a construction method of AIFV code trees based on integer programming (IP) for AIFV codes with two code trees. Although the IP problem is generally NP hard, the IP is used to solve more practical problems as the hardware of computers and the software of IP solvers develop. 

Before we treat AIFV code trees, we first consider the case of  binary Huffman code trees.
Let ${\cal X}=\{a_1, a_2, \cdots, a_{|{\cal X}|}\}$, $p_t=P_X(a_t)$, and $d_t=l(a_t)$. Then, 
the problem to obtain the binary Huffman code tree is 
equivalent to obtain $\{d_t\}$ that minimizes $\sum^{|{\cal X}|}_{t=1} p_t d_t$ under 
the Kraft inequality
\begin{align}
\sum^{|{\cal X}|}_{t=1} 2^{-d_t} \leq 1. \label{Kraft}
\end{align}
In this case, the inequality `$\leq$' in  \eqref{Kraft} can be replaced with equality `$=$'
because the optimal  $\{d_t\}$ always satisfies the equality in \eqref{Kraft}.

 In order to formalize this optimization problem as a 0-1 IP problem,
 we introduce binary variables $u_{t,d}$ such that 
$u_{t,d}=1$ if source symbol $a_t$ is assigned to a leaf of depth $d$ in a code tree,
and $u_{t,d}=0$ otherwise. 
Then, the optimization problem can be formalized as follows.
\begin{ip}\label{ip-1}
\begin{align}
\mbox{minimize} \quad&\sum_{t=1}^{|{\cal X}|}\ \sum_{d = 1}^{D}   u_{t,d}\, p_t \,d  \label{eq3-1} \\
\mbox{subject to} \quad&\sum_{t=1}^{|{\cal X}|}\ \sum_{d = 1}^{D} 2^{-d} u_{t,d}=1,\label{eq3-2} \\
&\sum_{d=1}^{D} u_{t,d}= 1, \qquad t = 1,2,\dots ,|{\cal X}|, \label{eq3-3} 
\end{align}
where $D$ is a positive integer constant, which represents the maximum depth considered in the IP problem.  
\end{ip}

\vspace{0.2cm}
Condition \eqref{eq3-3} guarantees that each $a_t$ is assigned to only one $d$,
and $d_t$ is determined as $d_t=d$ for $u_{t,d}=1$.
$D$ must be sufficiently large. But, large $D$ consumes computational time and memory. 
In many cases, it is sufficient that $D$ is several times as large as $\log_2 |{\cal X}|$.

\subsection{IP problem for binary AIFV code trees}\label{IP-bAIFV}
In order to obtain the optimal binary AIFV code for a given probability distribution $\{p_t\}$, we need to construct an IP problem that minimizes $L_{AIFV}=Q(T_0)L_0+Q(T_1)L_1$. 
However,  in such IP problems, we need a lot of variables because we must treat two code trees at once. Furthermore, since $Q(T_0)L_0$ and $Q(T_1)L_1$ include nonlinear terms, many additional variables and conditions are required to linearize nonlinear terms. 
Hence, although we can formalize an IP problem to obtain the global optimal solution, it becomes impractical or can treat only a small size of ${\cal X}$.
Therefore, in this subsection, we derive individual IP problems for $T_0$ and $T_1$ that can 
attain near-optimal $L_{AIFV}$, and we show in Section \ref{GP} that 
the global optimal AIFV code can be obtained by solving the individual IP problems finite times.

Since we can assign source symbols to master nodes in addition to leaves in the case of binary AIFV code, we introduce binary variables $v_{t,d}$, in addition to $u_{t,d}$, such that
 $v_{t,d}=1$ if source symbol $a_t$ is assigned to a master node of depth $d$,
and $v_{t,d}=0$ otherwise.
Then,  an IP problem to construct $T_0$ can be formalized 
as follows.
\begin{ip}\label{ip-2}
\begin{align}
\mbox{minimize} \quad&\sum_{t=1}^{|{\cal X}|}\ 
\sum_{d = 0}^{D}   p_t \left(u_{t,d}\,d+v_{t,d}\left(d+C_2\right)\right) \label{eq3-1-1} \\
\mbox{subject to} \quad&\sum_{t=1}^{|{\cal X}|}\ 
\sum_{d = 0}^{D} 2^{-d} \left(u_{t,d}+\frac{3}{4} v_{t,d}\right)=1,\label{eq3-2-1} \\
&\sum_{d=0}^{D} (u_{t,d}+v_{t,d})= 1, \hspace{6.8cm} t= 1,2,\dots ,|{\cal X}|, \label{eq3-3-1} \\
&\sum_{t=1}^{|{\cal X}|} \left( v_{t,d}+\frac{1}{2}v_{t,d+1}\right) 
-\sum_{\ell =d+2}^{D} \sum_{t=1}^{|{\cal X}|} 2^{d+2-\ell} \left( u_{t,\ell}+\frac{3}{4}v_{t,\ell} \right)  \leq 0, 
\qquad d = 0,1,\cdots ,D-2,\quad \label{eq3-4-1} 
\end{align}
where $C_2=2-\log_23 \approx 0.405$.
\end{ip}

Furthermore, an IP problem to derive $T_1$ is obtained by setting $u_{t,0}=v_{t,0}=0$ 
for all $t$ (or removing the case of $d=0$ in \eqref{eq3-1-1}--\eqref{eq3-4-1}) 
and replacing \eqref{eq3-2-1} 
with the following condition:
\begin{align}
\sum_{i=1}^{|{\cal X}|}\ 
\sum_{d = 1}^{D} 2^{-d} \left(u_{i,d}+\frac{3}{4} v_{i,d}\right)=\frac{3}{4}. \label{eq3-2-2} 
\end{align}

Condition \eqref{eq3-2-1} comes from \eqref{eq-20}, and condition \eqref{eq3-3-1} guarantees that each $a_t$ is assigned to only one of either leaves or  master nodes. 
The code trees are obtained by assigning $a_t$ to a leaf (resp.~a master node) of depth $d$ if the solution has $u_{t,d}=1$ (resp.~$v_{t,d}=1$). 

Note that $C_2$ in \eqref{eq3-1-1} and Eq.~\eqref{eq3-4-1} are newly introduced in IP problem \ref{ip-2} compared with IP problem \ref{ip-1}. We first consider why $C_2$ is required.

A leaf of depth $d$ has weight $2^{-d}$ in \eqref{eq3-2-1} 
while a master node of depth $d$ has weight $(3/4)2^{-d}$. 
Hence,  average code lengths $L_0$ and $L_1$ can be decreased 
by making many master nodes in $T_0$ and $T_1$, respectively.
On the other hand, this increases $P({\cal N}_1^{(0)})$ and $P({\cal N}_1^{(1)})$, 
and hence $Q(T_1)$  because of
 $Q(T_1|T_0) = P({\cal N}_1^{(0)})$ and  $Q(T_1|T_1) = P({\cal N}_1^{(1)})$. 
 Note that the global average code length is given by $L_{AIFV}=Q(T_0)L_0+Q(T_1)L_1$, and
 $L_1$ is much larger than $L_0$ because the root of $T_1$ cannot have a grandchild 
 with code symbols `00'. 
Therefore, $L_{AIFV}$ is not always minimized even if $L_0$ and $L_1$ are minimized individually.

Note that if a master node is used to encode a source symbol, we must use $T_1$, instead of $T_0$,  to encode the next source symbol. This means that master nodes have the cost $L'_{AIFV}-L_{AIFV}$ compared with leaves, where $L'_{AIFV}$ is the average code length of the case that we start
 the encoding with $T_1$ instead of $T_0$. 

Since we derive the code trees $T_0$ and $T_1$ by solving separate IP problems,  
it is hard to embed the exact cost into each IP problem. 
But, the optimal code trees have a good property such that every child of a node has approximately half probability weight of its parent node. So, as an approximation of exact cost, 
we can use the cost of the ideal case such that every node has two children with equal probability weight. 
In this case, the cost is given by $C_2=2-\log_23$ because 
the root of $T_0$ can have four grandchildren while the root of $T_1$ can have only three grandchildren.
Therefore,  cost $C_2$ is added for master nodes in \eqref{eq3-1-1}.

Next we consider \eqref{eq3-4-1}. This comes from Condition \ref{cond-2} shown in Section \ref{Kraft-bAIFV}. Each master node of depth $d$ requires a slave node of depth $d+1$ and a node or leaf of depth $d+2$.
Therefore, we cannot make master nodes of depth $d$ if there are not sufficient number 
of nodes or leaves at depth $d+2$.  
Let $N^{m}_d$ and $N^{nl}_{d+2}$ be the number of master node of depth $d$
and the number of nodes and leaves of depth $d+2$, respectively.
Then, $N^{m}_d$ is given by  
\begin{align}
N^{m}_d=\sum_{t=1}^{|{\cal X}|}  v_{t,d}. \label{eq3-4-2}
\end{align}
On the other hand, 
we can know the number of nodes and leaves of depth $d+2$
by calculating the Kraft's weight at depth $d+2$. Hence,
$N^{nl}_{d+2}$ is given by
\begin{align}
N^{nl}_{d+2}=\sum_{\ell =d+2}^{D} \sum_{t=1}^{|{\cal X}|} 2^{d+2-\ell} \left( u_{t,\ell}+\frac{3}{4}v_{t,\ell} \right).
\label{eq3-4-3}
\end{align}
Furthermore, there are $N^{m}_{d+1}$ master nodes of depth $d+1$, each of which 
requires one node or leaf of depth $d+3$. Since a node or leaf of depth $d+3$ has weight $2^{-1}$ at depth $d+2$, we must use $2^{-1}N^{m}_{d+1}$ out of $N^{nl}_{d+2}$
for master nodes of depth $d+1$. This means that the remaining $N^{nl}_{d+2}-2^{-1}N^{m}_{d+1}$ nodes and leaves of depth $d+2$ can be used for $N^{m}_d$ master nodes of depth $d$.
Hence, the condition \eqref{eq3-4-1} is required.

\subsection{IP problem for ternary AIFV code trees}\label{IP-tAIFV}
In order to obtain near-optimal ternary AIFV code,
we can formalize an IP problem for ternary AIFV code trees in the same way as binary AIFV code trees.

\begin{ip}\label{ip-3}
\begin{align}
\hspace{-0.2cm}\mbox{minimize} \hspace{0.1cm}
&\sum_{i=1}^{|{\cal X}|}\ \sum_{d = 0}^{D} p_i\left(u_{i,d}d+v_{i,d}\left(d+C_3\right)\right) \label{eq:3_objective} \\
\hspace{-0.2cm}\mbox{subject to} \hspace{0.1cm}
&\sum_{d = 0}^{D}  3^{-d} 
\left( z_d +\sum_{t=1}^{|{\cal X}|} \left(u_{t,d}+\frac{2}{3}v_{t,d} \right) \right)
 =1 \label{eq:3_T0}\\
&\sum_{d=0}^{\ell_{\max}} \left(u_{t,d}+v_{t,d}\right) = 1, \hspace{5.8cm} t = 1,2,\dots ,|{\cal X}|, \label{eq:3_00} \\
&\sum_{t=1}^{|{\cal X}|} v_{t,d} -\sum_{\ell =d+1}^{D}3^{d+1-\ell} \biggl( z_\ell 
 + \sum_{t=1}^{|{\cal X}|}\left(u_{t,\ell}+\frac{2}{3}v_{t,\ell}\biggr)
\right) \leq 0, 
\qquad d= 0,1,\cdots ,D-1. \label{eq:3_continuous} \end{align}
where $C_3=1-\log_32 \approx 0.369$.
\end{ip}

Furthermore, an IP problem to derive $T_1$  is obtained by 
setting $u_{t,0}=v_{t,0}=0$ for all $t$ (or removing the case of $d=0$ in \eqref{eq:3_objective}--\eqref{eq:3_continuous}) 
and replacing \eqref{eq:3_T0} with 
the following condition:
\begin{align}
\sum_{d = 1}^{D}  3^{-d} 
\left( z_d +\sum_{t=1}^{|{\cal X}|} \left(u_{t,d}+\frac{2}{3}v_{t,d} \right) \right)
 =\frac{2}{3}. \label{eq:3_T0-1}
\end{align}

The cost $C_3$ for incomplete internal nodes is given by $L'_{AIFV}-L_{AIFV}$ in 
the ideal case such that  every child of each node has equal probability weight.
Since the roots of $T_0$ and $T_1$ can have three and two children, respectively, 
in the ternary case, we have $C_3= \log_33-\log_32$.

Condition \eqref{eq:3_continuous} is required from Condition \ref{cond-1} shown in 
Section \ref{Kraft-KAIFV}, and it can be
derived in the same way as \eqref{eq3-4-1}.
But, since slave nodes do not exist in the ternary case, we do not need 
$\frac{1}{2}v_{t, d+1}$ in the first term of \eqref{eq3-4-1}.

A new binary variable $z_d$ is introduced in IP problem \ref{ip-3} compared with  IP problem \ref{ip-2}.
Note that the ternary Huffman code has one incomplete node in the code tree when $|{\cal X}|$ is even. Similarly  a ternary AIFV code may have one incomplete node in $T_0$ and/or $T_1$, 
which is not assigned any source symbol. Variable $z_d$ represents the pruned leaf of such an incomplete node. $z_d=1$ if there is the pruned leaf at level $d$, and $z_d=0$ otherwise.

We can represent the condition \eqref{eq:3_T0} without using $z_d$ as follows.
\begin{align}
\sum_{d = 0}^{D}   
\sum_{t=1}^{|{\cal X}|} 3^{-d}\left(u_{t,d}+\frac{2}{3}v_{t,d} \right) 
\leq 1. \label{eq:3_T0-2}
\end{align}
But, since the condition \eqref{eq:3_continuous} cannot be represented without $z_d$, 
 \eqref{eq:3_T0} is used rather than \eqref{eq:3_T0-2}.
Since the pruned leaf must have the longest depth if it exists, 
we have $z_{\hat{d}}=1$ for $\hat{d}\equiv \max\{d : u_{t,d}=1, t=1,2, \cdots |{\cal X}|\}$  and $z_{\hat{d}}=0$ for $d\neq \hat{d}$ in the optimal $T_0$ and $T_1$. 
But these conditions are not explicitly included in IP problem \ref{ip-3} because the optimal code trees can be obtained without these conditions.

\begin{remark}
IP problem \ref{ip-3} can be applied to the $K$-ary AIFV codes with two code trees $T_0$ and $T_{K-j}$
explained in Remark \ref{remark-1} by modifying 2, 3, $C_3$ and $z_d$ in 
\eqref{eq:3_objective}-\eqref{eq:3_continuous} as follows:
\begin{align*}
& 3\rightarrow K, \qquad 2 \rightarrow K-j,  \qquad C_3 \rightarrow C_{K,j}=1-\log_K (K-j),\\
& z_d \in\{0,1\} \rightarrow z_d \in\{0,1, \cdots, K-2\}. 
\end{align*}
We can also construct IP problems for general $K$-ary AIFV code trees
by using binary variables $v_{t,d}^{(j)}$ to represent incomplete internal nodes with $j$ children for $1\leq j \leq K-2$ instead of $v_{t,d}$ used in IP problem \ref{ip-3}. 
But,  the necessary number of variables increases and each condition described in `subjet to'
becomes long as $K$ becomes large. Therefore, it is hard to treat large $K$ practically because of
time and/or space complexity.
\end{remark}

\subsection{Global Optimaization} \label{GP}
In IP problems \ref{ip-2} and \ref{ip-3}, costs $C_2$ and $C_3$ are determined 
based on the ideal code trees such that every child of each node has equal probability weight. But, since the code trees $T_0$ and $T_1$ obtained by IP Problem \ref{ip-2} (or \ref{ip-3}) do not attain 
the perfect balance of probability weight, they are not the optimal AIFV code trees generally. 
So, we calculate new cost $C$ based on the obtained code trees $T_0$ and $T_1$, and we derive new code trees for the new cost by solving again IP Problem \ref{ip-2} (or \ref{ip-3}). In this section, we show that the global optimal code trees can be obtained by repeating this procedure.

Let $C^{(m-1)}$ is the $(m-1)$-th cost and let $T_0^{(m)}$ and $T_1^{(m)}$ be the $m$-th AIFV code trees obtained by solving the IP problem for cost $C^{(m-1)}$.  $C^{(0)}$ is the initial cost.
Furthermore, 
let $L_0^{(m)}$ and $L_1^{(m)}$ be the average code length of $T_0^{(m)}$ and $T_1^{(m)}$,
respectively, and let $q_0^{(m)}$ and $q_1^{(m)}$ be the transition probabilities of code trees $T_0^{(m)}$ and $T_1^{(m)}$, which are defined by $q_0^{(m)}\equiv Q(T_1^{(m)}|T_0^{(m)})=P(\mbox{${\cal N}_1^{(0)}$ in $T_0^{(m)}$})$
and $q_1^{(m)}\equiv Q(T_0|T_1)=P(\mbox{${\cal N}_0^{(1)}$ in $T_1^{(m)}$})$.

Then, we consider the following algorithm. 
\begin{algorithm} \label{alg-1}
\begin{description}
\item{1.} Set $m=1$ and $C^{(0)}=C$ for given initial cost $C$.
\item{2.} Obtain $T_0^{(m)}$ and $T_1^{(m)}$ by solving IP problem \ref{ip-2} (or \ref{ip-3}) 
for cost $C^{(m-1)}$.
\item{3.} Calculate  $(L_0^{(m)}, q_0^{(m)})$ for $T_0^{(m)}$ and  $(L_1^{(m)}, q_1^{(m)})$ for $T_1^{(m)}$.
\item{4.} Update cost as follows.
\begin{align}
C^{(m)}=\frac{L_1^{(m)} -L_0^{(m)}}{q_0^{(m)}+q_1^{(m)}} \label{eq4-5}
\end{align}
\item{5.} If $C^{(m)}=C^{(m-1)}$, then exit. Otherwise, increment $m$ and go to step 2.
\end{description}
\end{algorithm}

\vspace{0.2cm}
We can use any $C$ for the initial cost. But, if we use $C_2=2-\log_23$ and $C_3=1-\log_32$ 
as the initial cost in the binary and ternary cases, respectively, $T_0^{(1)}$ and $T_1^{(1)}$ become near-optimal code trees.

\vspace{0.2cm}
\begin{theorem}
The binary AIFV code and the ternary AIFV code obtained by Algorithm \ref{alg-1} are optimal.
\end{theorem}
\vspace{0.2cm}

{\it Proof} \quad
We first prove that  Algorithm \ref{alg-1} stops after finite iterations.
First note that for $T_0^{(m)}$, the objective function \eqref{eq3-1-1}  in IP problem \ref{ip-2}
(or \eqref{eq:3_objective} in IP problem \ref{ip-3} )  can be represented as 
\begin{align}
L_0^{(m)}+C^{(m-1)} q_0^{(m)}. \label{eq4-1}
\end{align}
Similarly, the object function for $T_1^{(m)}$ can be represented as 
\begin{align}
L_1^{(m)}+C^{(m-1)} (1-q_1^{(m)}). \label{eq4-2}
\end{align}
Since $C^{(m-1)}$ is fixed in the IP problem used in step 2 of Algorithm \ref{alg-1}, 
the minimization of \eqref{eq4-2} is equivalent to the minimization of 
\begin{align}
L_1^{(m)} - C^{(m-1)} q_1^{(m)}. \label{eq4-3}
\end{align}
On the other hand, the global average code length $L_{AIFV}^{(m)}$ for 
$T_0^{(m)}$ and $T_1^{(m)}$ is given by
\begin{align}
L_{AIFV}^{(m)}=\frac{q_1^{(m)} L_0^{(m)}+ q_0^{(m)}L_1^{(m)}}{q_0^{(m)}+q_1^{(m)}}.
\end{align}

Since $T_0^{(m)}$ and $T_1^{(m)}$ are optimal trees that minimize \eqref{eq4-1}
and \eqref{eq4-3} for $C^{(m-1)}$, 
the following inequalities hold for any code trees $T_0$ with $(L_0, q_0)$ and $T_1$
with $(L_1, q_1)$.
\begin{align}
L_0^{(m)}+C^{(m-1)} q_0^{(m)}\leq L_0 +C^{(m-1)}q_0, \label{eq4-4}\\
L_1^{(m)}-C^{(m-1)} q_1^{(m)}\leq L_1 -C^{(m-1)}q_1. \label{eq4-4-1}
\end{align}
Hence if we substitute $T_0=T_0^{(m-1)}$ and $T_1=T_1^{(m-1)}$ into 
\eqref{eq4-4} and \eqref{eq4-4-1}, respectively, we have the following inequalities.
\begin{align}
L_0^{(m)}+C^{(m-1)} q_0^{(m)}
&\leq L_0^{(m-1)}+C^{(m-1)}q_0^{(m-1)} \nonumber\\
& =L_{AIFV}^{(m-1)}  \label{eq4-6}\\
L_1^{(m)}-C^{(m-1)} q_1^{(m)}
&\leq L_1^{(m-1)} -C^{(m-1)}q_1^{(m-1)} \nonumber\\
& = L_{AIFV}^{(m-1)} \label{eq4-7}
\end{align}
If $C^{(m)}< C^{(m-1)}$, we obtain from \eqref{eq4-6} that
\begin{align}
L_{AIFV}^{(m)}&=L_0^{(m)}+C^{(m)} q_0^{(m)} \nonumber\\
&< L_0^{(m)}+C^{(m-1)} q_0^{(m)}\nonumber\\
&\leq L_{AIFV}^{(m-1)} \label{eq4-8}.
\end{align}
Similarly, if $C^{(m)}> C^{(m-1)}$, we have from \eqref{eq4-7} that
\begin{align}
L_{AIFV}^{(m)}&=L_1^{(m)}-C^{(m)} q_1^{(m)} \nonumber\\
&< L_1^{(m)}-C^{(m-1)} q_1^{(m)}\nonumber\\
&\leq L_{AIFV}^{(m-1)} \label{eq4-9}.
\end{align}
Therefore, if $C^{(m)}\neq C^{(m-1)}$,  we have that $L_{AIFV}^{(m)}< L_{AIFV}^{(m-1)}$.
Since $L_{AIFV}^{(m)}>0$ for any $m$, we can conclude that $L_{AIFV}^{(m)}$ converges as $m\rightarrow \infty$. Furthermore, since the number of code trees is finite, the convergence is 
achieved with finite $m$, i.e. $C^{(m)}= C^{(m-1)}$ occurs and Algorithm 1 stops after finite iterations.

Next we prove that the obtained AIFV code trees are optimal when Algorithm 1 stops.
Assume that Algorithm 1 stops at $m=\hat{m}$, and  $T_0^{(\hat{m})}$ and  $T_1^{(\hat{m})}$ are the obtained AIFV code trees that satisfy $C^{(\hat{m})}=C^{(\hat{m}-1)}$.
If this pair $(T_0^{(\hat{m})}, T_1^{(\hat{m})})$ is not globally optimal, there exists the optimal pair of code trees $(T_0^*,  T_1^*)$ with $(L_0^*, L_1^*, q_0^*, q_1^*)$ such that
\begin{align}
 L_{AIFV}^{(\hat{m})}> L^*_{AIFV}. \label{eq4-10}
\end{align}
Then, we have for $C^*\equiv (L_1^*-L_0^*)/(q_0^*+q_1^*)$ that
\begin{align}
L^*_{AIFV}= L_0^* +C^* q_0^*  = L_1^* - C^* q_1^*.  \label{eq4-11}
\end{align}
Hence, if $C^*\geq C^{(\hat{m})}$, we have
\begin{align}
L_{AIFV}^{(\hat{m})}&=L_0^{(\hat{m})}+ C^{(\hat{m})} q_0^{(\hat{m})} \nonumber\\
&= L_0^{(\hat{m})}+ C^{(\hat{m}-1)} q_0^{(\hat{m})}\nonumber\\
& \leq L_0^*+ C^{(\hat{m}-1)}q_0^* \nonumber\\
& \leq  L_0^*+ C^*q_0^*\nonumber\\
& =L^*_{AIFV}, \label{eq4-13}
\end{align}
where the first inequality and the last equality hold from \eqref{eq4-4} and \eqref{eq4-11},
respectively.
Similarly if $C^*\leq C^{(\hat{m})}$, we have
\begin{align}
L_{AIFV}^{(\hat{m})}&=L_1^{(\hat{m})}- C^{(\hat{m})} q_1^{(\hat{m})} \nonumber\\
&= L_1^{(\hat{m})}- C^{(\hat{m}-1)} q_1^{(\hat{m})}\nonumber\\
& \leq L_1^* - C^{(\hat{m}-1)}q_1^* \nonumber\\
& \leq  L_1^*-  C^*q_1^*\nonumber\\
& =L^*_{AIFV}. \label{eq4-14}
\end{align}
Since \eqref{eq4-13} and \eqref{eq4-14} contradict with \eqref{eq4-10}, 
the pair of obtained code trees $(T_0^{(\hat{m})}, T_1^{(\hat{m})})$ must be globally optimal.
 \begin{flushright}Q.E.D.\end{flushright}

\section{Performance of binary and ternary AIFV codes}\label{sec-5}
In this section, we compare numerically the performance of AIFV codes with Huffman codes.  
For ${\cal X}=\{a_1, a_2 \cdots, a_{|{\cal X}|}\}$, 
we consider the following three kinds of source distributions:
\begin{align}
P_X^{(0)}(a_t)&=\frac{1}{|{\cal X}|},  \label{eq5-1}\\
P_X^{(1)}(a_t) &=\frac{t}{A_1},        \label{eq5-2} \\
P_X^{(2)}(a_t)&=\frac{t^2}{A_2},   \label{eq5-3}
\end{align}
where $A_1=\sum_{t=1}^{|{\cal X}|} t$ and $A_2=\sum_{t=1}^{|{\cal X}|} t^2$ are normalizing constants.

\begin{figure}[t]
  \begin{center}
    \includegraphics[clip, scale = 0.35]{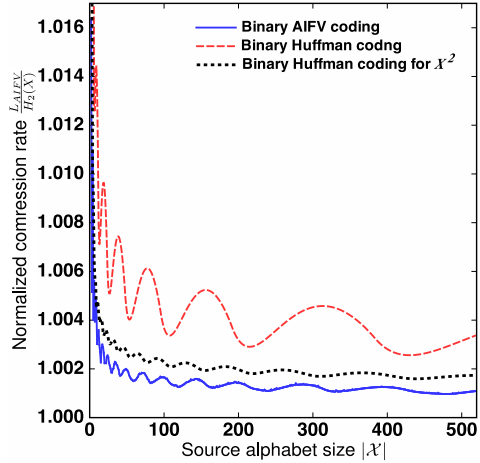}
    \caption{Comparison between binary AIFV coding and Huffman coding for $P_X^{(1)}$.}
     \label{fig10}
  \end{center}
\end{figure}

\begin{figure}[t]
  \begin{center}
    \includegraphics[clip, scale = 0.35]{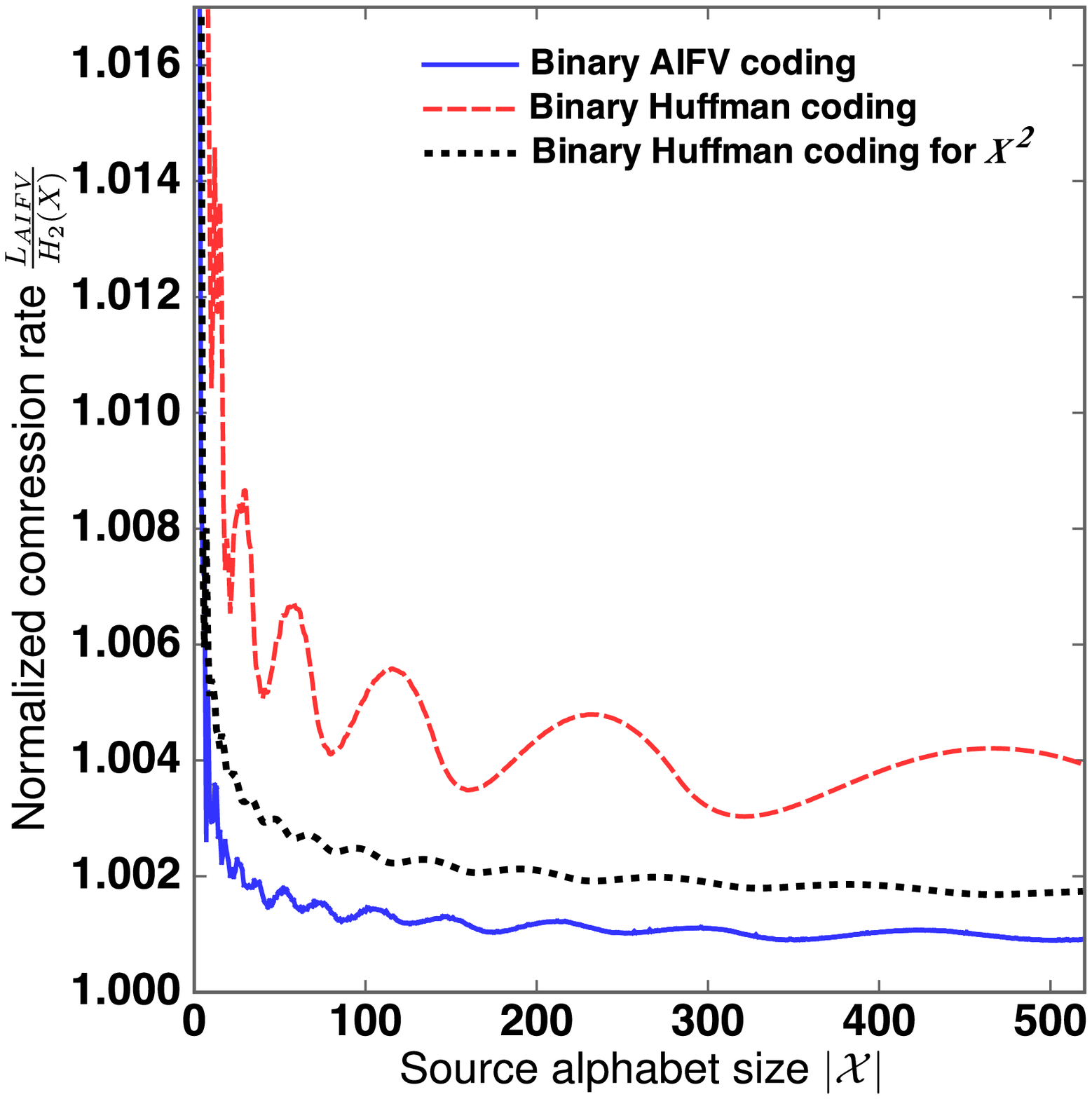}
    \caption{Comparison between binary AIFV coding and Huffman coding for $P_X^{(2)}$.}
     \label{fig11}
  \end{center}
\end{figure}

\begin{figure}[t]
  \begin{center}
    \includegraphics[clip, scale = 0.35]{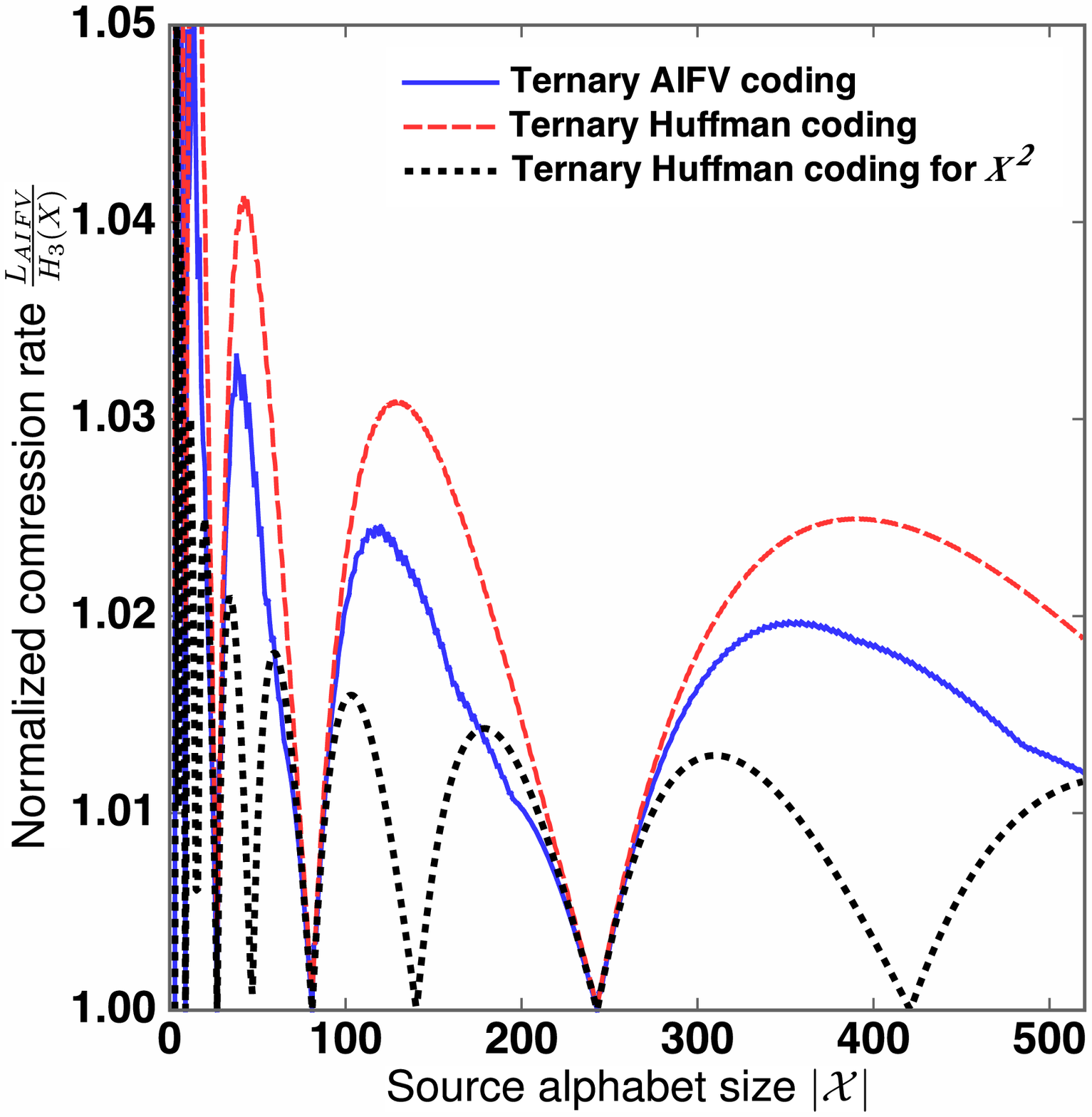}
    \caption{Comparison between ternary AIFV coding and Huffman coding for $P_X^{(0)}$}
    \label{fig12}
  \end{center}
\end{figure}

\begin{figure}[t]
  \begin{center}
    \includegraphics[clip, scale = 0.35]{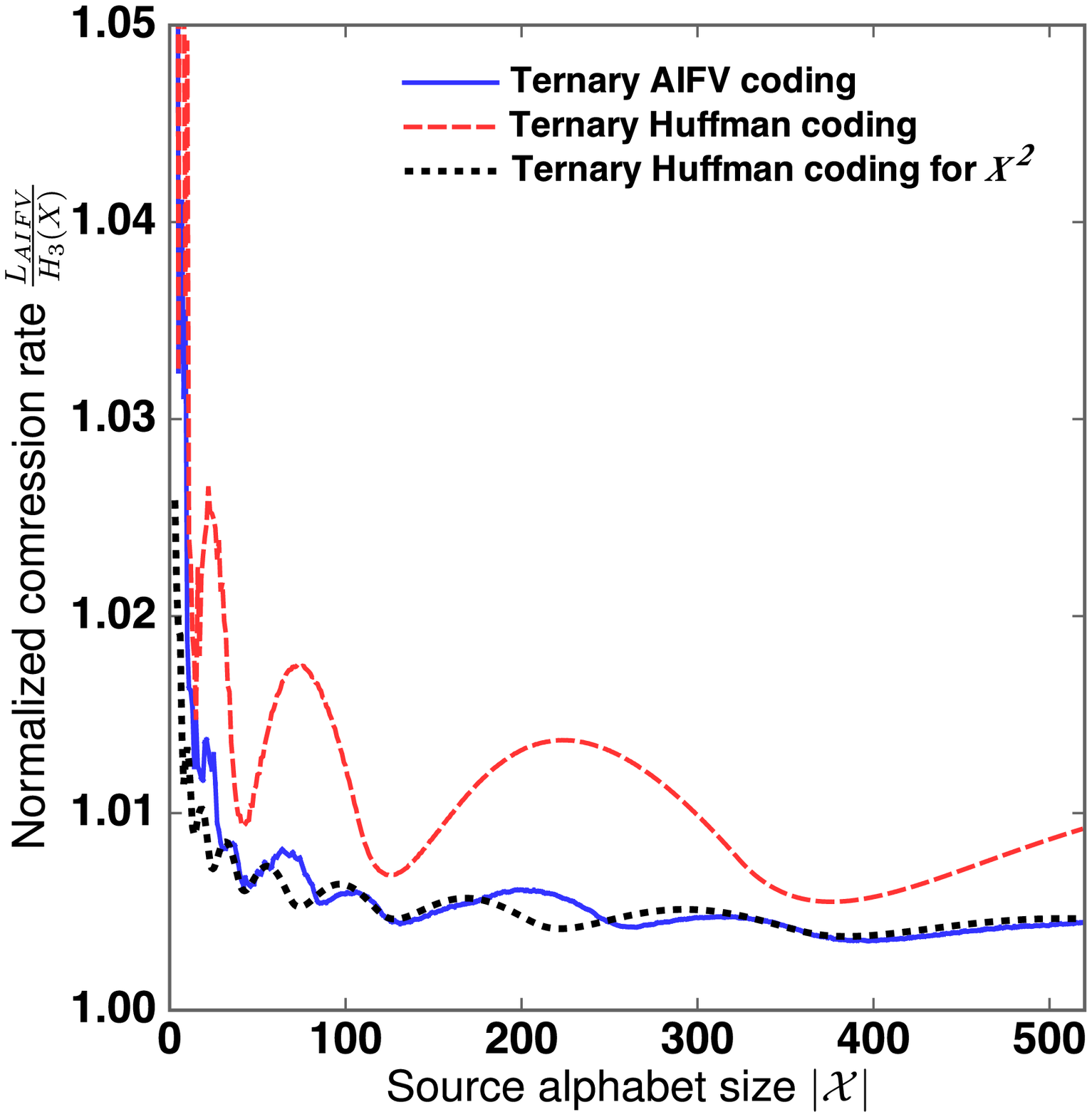}
    \caption{Comparison between ternary AIFV coding and Huffman coding for $P_X^{(1)}$.}
    \label{fig13}
  \end{center}
\end{figure}

\begin{figure}[t]
  \begin{center}
    \includegraphics[clip, scale = 0.35]{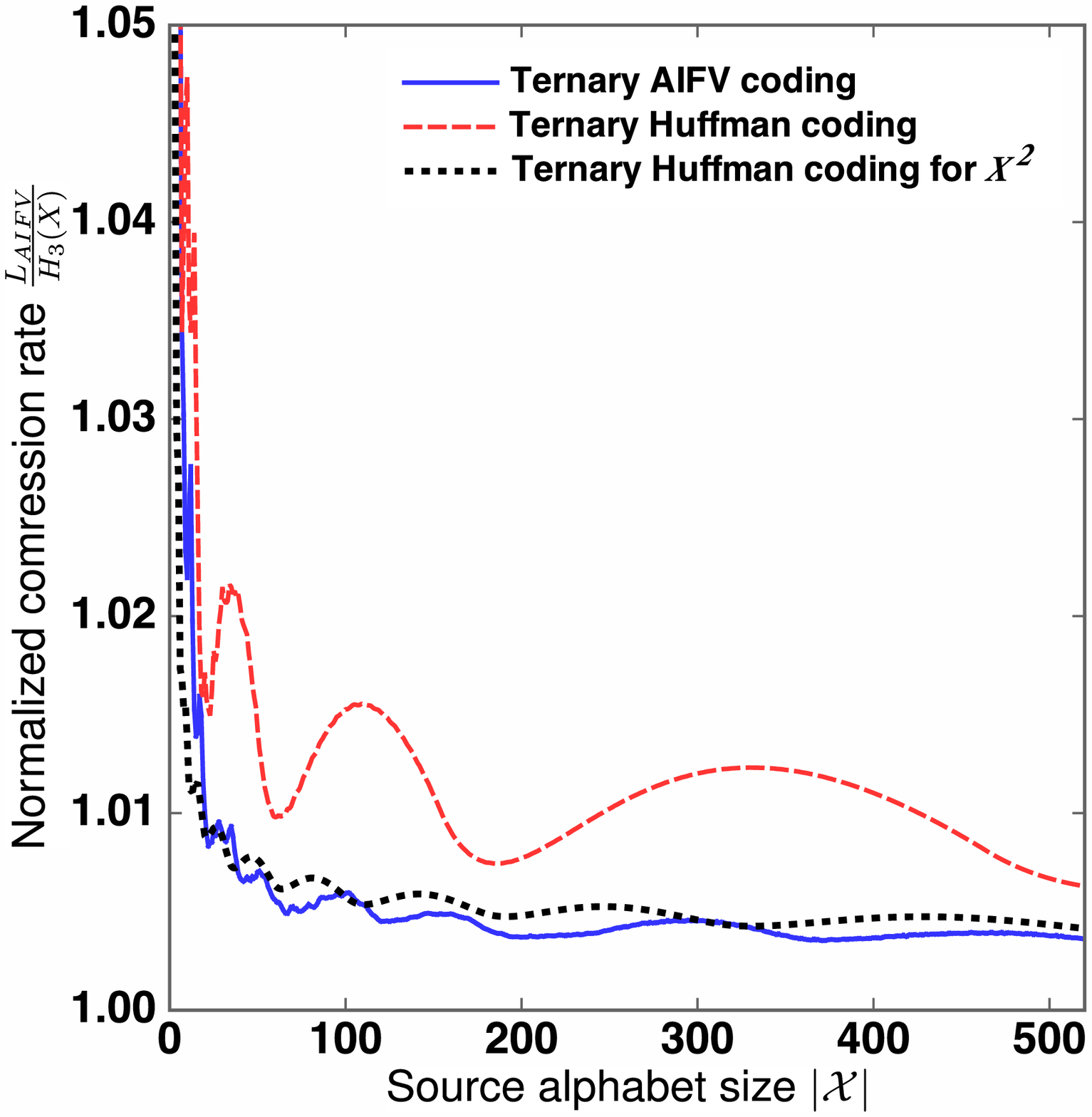}
    \caption{Comparison between ternary AIFV coding and Huffman coding for $P_X^{(2)}$.}
    \label{fig14}
  \end{center}
\end{figure}

The performance of AIFV codes is compared with 
Huffman codes and Huffman codes for ${\cal X}^2$ 
in Figs.~\ref{fig10}--\ref{fig11} (resp.~Figs.~\ref{fig12}--\ref{fig14})
for the binary (resp.~ternary) case\footnote{Figures 3--6 and 8 in \cite{Y-W-2013} are not correct
although the algorithms shown in \cite{Y-W-2013} are correct.}.

The comparison for  $P_X^{(0)}$   is omitted in the binary case because the compression rate of AIFV codes is equal to the one of Huffman codes. The AIFV codes are derived by Algorithm
\ref{alg-1}. 

In the figures, the vertical line represents the normalized compression rate defined by
$L_{AIFV}/H_2(X)$ and $L_H/H_2(X)$  (resp.~$L_{AIFV}/H_3(X)$ and $L_H/H_3(X)$) for the binary
(resp.~ternary) case. The horizontal line stands for the size of source alphabet. 
We note from Figs.~\ref{fig10}--\ref{fig14}
that the AIFV codes can attain better compression rate than the Huffman codes in all cases.
Furthermore, in the cases of $P_X^{(1)}$ and $P_X^{(2)}$,
the binary AIFV codes can beat even the Huffman codes for ${\cal X}^2$ and the ternary AIFV codes can attain almost the same compression rate as the Huffman codes for ${\cal X}^2$.



The Huffman coding for ${\cal X}^2$ has demerits such that 
the size of Huffman code tree increases to roughly $|{\cal X}|^2$, and the encoding and decoding delay of the first source symbol of $(x_1, x_2) \in {\cal X}^2$ becomes large as $|{\cal X}|$ becomes large. 
On the other hand, in AIFV coding, the size of code trees 
is roughly $2|{\cal X}|$ for these binary and ternary cases\footnote{In the $K$-ary case for $K\geq 3$, the size of AIFV code trees is roughly $(K-1)|{\cal X}|$.}, and encoding delay is zero and decoding delay is at most two bits 
(resp.~one code symbol) in binary (resp.~$K$-ary for $K\geq 3$) case.
Hence, from the viewpoints of coding delay and memory size,  AIFV coding is superior to  Huffman coding for ${\cal X}^2$ when $|{\cal X}|$  is large.

Finally we remark that 
if we use $C=C_2=2-\log_23$ (resp.~$C=C_3=1-\log_32$) as the initial cost
in Algorithm \ref{alg-1} for the binary (resp.~ternary) case, 
$L_{AIFV}^{(1)}$ is often optimal without iteration. Furthermore, even if $L_{AIFV}^{(1)}$ is not optimal, the improvement by the iteration of Algorithm \ref{alg-1} is within only 0.1\% compared with $L_{AIFV}^{(1)}$ in all the cases of $P_X^{(0)}$, $P_X^{(1)}$, and $P_X^{(2)}$. 
This means that if we use $C=C_2$ (resp.~$C=C_3$) in IP problem \ref{ip-2} (resp.~IP problem \ref{ip-3}),
we can obtain the optimal or near-optimal AIFV codes by solving the IP problem for $T_0$ and $T_1$ only once without using Algorithm \ref{alg-1}. 

\section{Conclusion}\label{sec-6}
In this paper, we proposed binary and $K$-ary (for $K\geq 3$) AIFV coding for stationary memoryless sources,
and we showed that the optimal AIFV codes can be obtained by solving integer programing problems for the binary and ternary cases.
Furthermore, by calculating the compression rate numerically for several source distributions, 
we clarified that the AIFV coding can beat Huffman coding.

The following are open problems:  obtain a tight upper bound of $L_{AIFV}$ given in \eqref{eq-16}, 
obtain a simple algorithm to derive the optimal binary AIFV codes and/or the optimal $K$-ary AIFV codes.

The AIFV codes proposed in this paper are devised such that decoding delay is at most one code symbol (resp.~two bits) in $K$-ary (resp.~binary) case. But, if decoding delay is allowed  more than one code symbol (resp.~two bits), it may be possible to construct non-instantaneous FV codes that can attain better compression rate than the AIFV codes. It is also an interesting open problem to obtain the best non-instantaneous FV codes for a given maximum decoding delay.

\ifCLASSOPTIONcompsoc
\else
\fi

\ifCLASSOPTIONcaptionsoff
  \newpage
\fi

\end{document}